\documentclass[12pt]{article}
\usepackage{amsthm,amsmath,amsfonts,amssymb}
\usepackage[authoryear,round]{natbib}
\usepackage[colorlinks,citecolor=blue,urlcolor=blue]{hyperref}
\usepackage{graphicx}
\usepackage{color}
\usepackage[dvipsnames]{xcolor}
\usepackage{amsmath}
\usepackage{bbm}
\usepackage{float}
\usepackage{setspace}
\usepackage[margin=1in]{geometry}
\usepackage{authblk}
\usepackage[ruled,vlined]{algorithm2e}
\usepackage{pdflscape}

\numberwithin{equation}{section}

\newcommand{\NN}{\mathbb{N}}

\newcommand{\RR}{\mathbb{R}}

\newcommand{\RC}[2]{\mathrm{RC}_{#2}(#1)}

\DeclareMathOperator*{\argmin}{arg\,min}

\newtheorem{property}{Property}[section]

\theoremstyle{remark}

\theoremstyle{definition}

\begin{document}



\title{\bf Modelling non-stationarity in asymptotically independent extremes}
\author[1,2*]{C. J. R. Murphy-Barltrop}
\author[3]{J. L. Wadsworth}
\affil[1]{Technische Universität Dresden, Institut Für Mathematische Stochastik, Helmholtzstraße 10, 01069 Dresden, Germany}
\affil[2]{Center for Scalable Data Analytics and Artificial Intelligence (ScaDS.AI) Dresden/Leipzig, Germany}
\affil[3]{Department of Mathematics and Statistics, Lancaster University LA1 4YF, United Kingdom}
\affil[*]{Correspondence to: callum.murphy-barltrop@tu-dresden.de}
\date{\today}

\maketitle

\singlespacing
\begin{abstract}
  In many practical applications, evaluating the joint impact of combinations of environmental variables is important for risk management and structural design analysis. When such variables are considered simultaneously, non-stationarity can exist within both the marginal distributions and dependence structure, resulting in complex data structures. In the context of extremes, few methods have been proposed for modelling trends in extremal dependence, even though capturing this feature is important for quantifying joint impact. Moreover, most proposed techniques are only applicable to data structures exhibiting asymptotic dependence. Motivated by observed dependence trends of data from the UK Climate Projections, we propose a novel semi-parametric modelling framework for bivariate extremal dependence structures. This framework allows us to capture a wide variety of dependence trends for data exhibiting asymptotic independence. When applied to the climate projection dataset, our model detects significant dependence trends in observations and, in combination with models for marginal non-stationarity, can be used to produce estimates of bivariate risk measures at future time points.
\end{abstract}
\noindent
{\it Keywords:}  Extremal Dependence, Non-stationary Processes, Multivariate Extremes


\doublespacing

\section{Introduction} \label{sec:intro}
Modelling joint tail behaviour of multivariate datasets is important in a wide variety of applications, including nuclear regulation \citep{OfficeforNuclearRegulation2018}, neuroscience \citep{Guerrero2021} and flood risk analysis \citep{Gouldby2017}. When analysing multivariate extremes, it is important to capture the dependence structure at extreme levels appropriately. In certain applications, one would expect the extremes to occur simultaneously -- a situation termed asymptotic dependence -- whilst in others, joint occurrence of the very largest events cannot happen -- a situation termed asymptotic independence. Section \ref{sec:background} explains these concepts in detail. The study of extremal dependence structures is well established, and a wide range of statistical modelling techniques have been proposed \citep{Coles1991,Ledford1997,Heffernan2004}. 

Extremal dependence between two variables may be summarised by bivariate risk measures. A variety of risk measures have been proposed in the literature \citep{Serinaldi2015}, and are selected according to the needs of an analysis. For this paper, we restrict attention to one particular measure known as the return curve due to its utilisation in a variety of practical applications \citep{Murphy-Barltrop2023}. Given a small probability $p$ and a random vector $(X,Y)$ with strictly continuous marginal distributions, the $p$-probability return curve is given by $\RC{p}{} := \{ (x,y) \in \RR^2 \mid \Pr(X>x,Y>y) = p \}$, with corresponding return period $1/p$. This curve directly extends the concept of a return level from the univariate framework \citep{Coles2001} to the bivariate setting. These curves, which provide a summary of joint tail behaviour, are widely used in practice to derive extremal conditions during the design analysis of many ocean and coastal structures, including oil rigs \citep{Jonathan2014}, railway lines \citep{Gouldby2017} and wind turbines \citep{Manuel2018}. 

However, in many real world scenarios, datasets exhibit non-stationarity; this feature can result in extremal dependence structures that are not fixed due to covariate influences on the underlying processes. In this setting, there is no longer a meaningful or fixed definition of a return curve. We therefore expand the definition of this risk measure to be covariate-dependent, resulting in a non-stationary counterpart; see \citet{Rootzen2013} and \citet{Serinaldi2015} for related discussion. Given some covariates $\mathbf{Z}_t = (Z_{1,t}, \hdots, Z_{g,t})$, $g \in \NN, \; t \in \{1,2,\ldots,n\}$, where $t$ denotes time, let $\{X_t,Y_t\}$ denote a conditionally stationary process, i.e., the distribution of $(X_t,Y_t) \mid \mathbf{Z}_t$ does not depend on $t$ \citep{Caires2005}. In this setting, we define the $p$-probability return curve at a covariate realisation $\mathbf{z}_t$ to be $\RC{p}{\mathbf{z}_t} := \{ (x,y) \in \RR^2 \mid \Pr(X_t>x,Y_t>y \mid \mathbf{Z}_t = \mathbf{z}_t) = p\}$. Evaluation of $\RC{p}{\mathbf{z}_t}$ over different values of $t$ allows one to explore joint extremal behaviour over time, and thus may be useful when designing ocean and coastal structures for future climates. 

In a practical setting, we wish to derive estimates of non-stationary return curves for environmental datasets to evaluate the changing risk with covariates. Our methodology is motivated particularly by non-stationarity observed in data obtained from the UK Climate Projections (UKCP18) under emissions scenario RCP 8.5. This corresponds to the `worst-case' scenario, whereby greenhouse gas emissions continue to rise throughout the 21st century \citep{MetOffice2018}. As such, data from these projections can be used as a risk management tool to help mitigate against the impacts of climate change in a conservative manner. Specifically, we focus in this work on relative humidity and temperature projections over the summer months (June, July and August) at a grid cell containing the UK's Heysham nuclear power station. Denoting relative humidity by $\mathrm{RH}_t \in [0,100]$ for $t \in \{1,2,\hdots,n\}$, we define a `dryness' variable as $\mathrm{Dr}_t := 100 - \mathrm{RH}_t \in [0,100]$. 

Data is only considered for summer months since extremal dependence structures vary significantly across meteorological seasons and worst extremes tend to occur in summer; see the Supplementary Material for further details. In the context of nuclear safety, both high temperature and high dryness (low humidity) values are independently identified as primary hazards by the UK's Office for Nuclear Regulation (ONR) \citep{OfficeforNuclearRegulation2018}. As a result, pre-set `design values' of either variable, corresponding to a `1 in 10,000-year' event, are used to inform the design bases of UK-based nuclear sites. Moreover, the combination of high temperature and high dryness values has been identified as a relevant safety consideration \citep{Knochenhauer2004,OfficeforNuclearRegulation2021}, since this combination is often associated with drought-like conditions. Such conditions have the potential for catastrophic consequences, including loss of foundation support to facilities and loss of water supply. Therefore, evaluating the joint extremal behaviour for this particular combination of hazards may provide useful information about joint risk over the observation period.

\begin{figure}[tb]
    \centering
    \includegraphics[width=.95\textwidth]{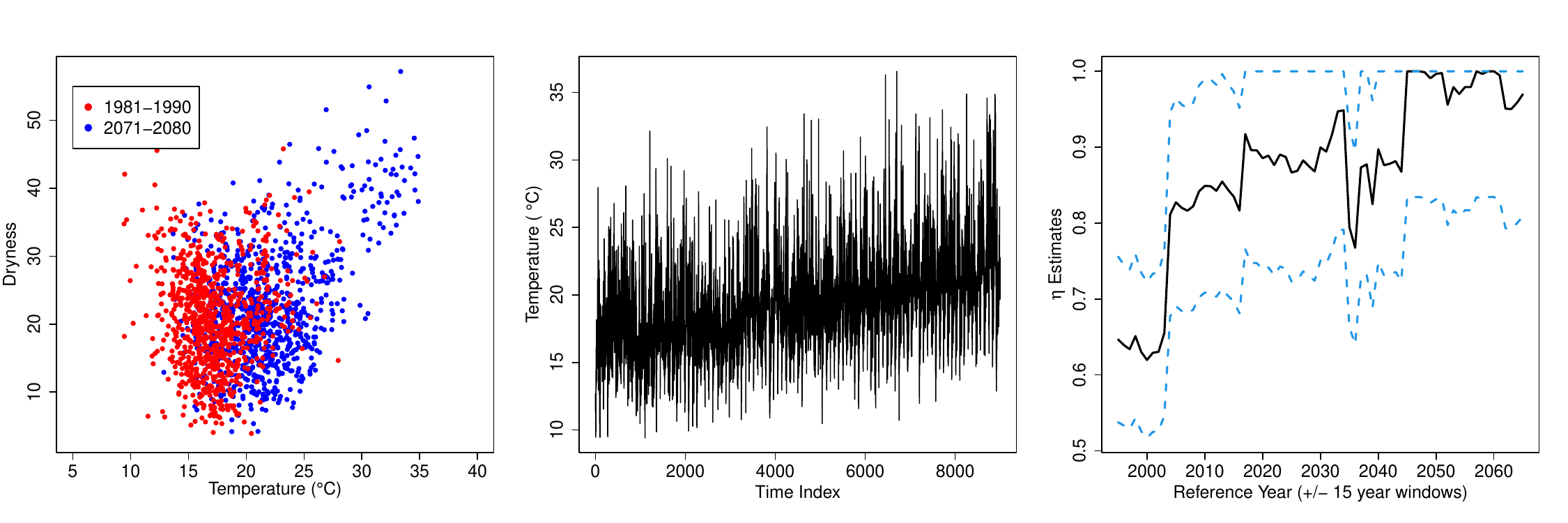}
    \caption{Left: Plot of first and last 10 years of combined projections, given in red and blue, respectively. Centre: Plot of Heysham temperature time series. Right: Plot of $\eta$ estimates over rolling windows (solid black lines), alongside $95\%$ pointwise confidence intervals (dotted blue lines).}
    \label{fig:UKCP18}
\end{figure}

The dataset of temperature and dryness at the start and end of the time period, along with the temperature time series, are plotted in the left and centre panels of Figure \ref{fig:UKCP18}, respectively. Clear non-stationary trends can be observed within both marginal data sets; these trends are likely a result of seasonal behaviour combined with long term trends due to climate change. 

To assess trends in dependence, we can calculate suitable coefficients using rolling windows of data, assuming local stationarity within each window. Rolling windows are defined by taking a reference year $y \in \{1995,1996,\dots,2065\}$ and considering all data for the months June, July and August within the interval $[y-15,y+15]$: this results in 2,790 observations for each window. The right panel of Figure \ref{fig:UKCP18} demonstrates a clear trend in an extremal dependence coefficient labelled $\eta$ \citep{Ledford1996}; this measure summarises the dependence between the most extreme observations, with larger values corresponding to a higher degree of positive dependence. Further discussion can be found in Section \ref{subsec:mvevt}. The illustrated trend suggests the probability of extreme observations occurring simultaneously is increasing over time, motivating the need for modelling techniques that can capture trends of this nature. We return to a detailed analysis of this dataset in Section \ref{sec:case_study}.



The majority of existing techniques for modelling multivariate extremes assume stationarity in the joint tail structure. Furthermore, of the approaches that can accommodate non-stationarity, most are suitable only for datasets exhibiting asymptotic dependence, as we discuss in Section \ref{subsec:ns_extreme_dep}. This is restrictive since in practice, asymptotic independence is often observed \citep{Ledford1996,Ledford1997}; this is further evidenced by estimated $\eta$ values for the UKCP18 dataset, which indicate the presence of asymptotic independence, at least throughout most of the observation period. 


We propose a new method for capturing non-stationary extremal dependence structures when asymptotic independence is present, based on a non-stationary extension to the \citet{Wadsworth2013} modelling framework. In doing so, we are able to evaluate and visualise trends across the entire extremal dependence structure. This is in contrast to other approaches, where implementation may be limited to trends in one-dimensional summary measures, such as the coefficient of tail dependence \citep{Ledford1996} or the extremal coefficient \citep{Frahm2006}.

This paper is structured as follows: Section \ref{sec:background} recalls existing methodology for capturing tail behaviour in the stationary and non-stationary settings for both univariate and multivariate random vectors. Section \ref{sec:novel_model} introduces a range of novel estimators for a quantity describing extremal dependence in the non-stationary setting. We also propose methodology for non-stationary return curve estimation using these estimators. Section \ref{sec:sim_study} details a simulation study, where we compare performance of these estimators. In Section \ref{sec:case_study}, we apply our model to the UKCP18 dataset. Our approach is able to reveal clear trends in the extremal dependence of this process, and estimates of return curves are obtained. We conclude in Section \ref{sec:discussion} with a discussion and outlook on future work. 



\section{Background} \label{sec:background}
\subsection{Univariate extreme value theory} \label{subsec:unievt}
In the univariate setting, one of the most popular techniques for capturing tail behaviour is known as the peaks-over-threshold approach, whereby a generalised Pareto distribution (GPD) is fitted to all exceedances of some high threshold. This is justified by the Pickands-Balkema-de Haan theorem \citep{Balkema1974,Pickands1975}, which states that for a random variable $X$ satisfying certain regularity conditions, there exists a normalising function $c(u) > 0$ such that 
\begin{equation} \label{eqn:gpd_cdf}
    \Pr \left(\frac{X - u}{c(u)} \leq x \;  \Big\vert \; X>u\right) \to G(x) := 1 - \left\{1 + \frac{\xi x}{\tau} \right\}_+^{-1/\xi}, \hspace{.5em} x > 0, \; (\tau,\xi) \in \RR^+ \times \RR,
\end{equation}
as $u \to x^F := \sup\{x : F(x) < 1 \}$; see also \citet{Coles2001}. Here, $G(x)$ is the cumulative distribution function of a GPD, with scale and shape parameters, $\tau$ and $\xi$, respectively, and $z_+ = \max(0,z)$. The shape parameter dictates the behaviour of the tail, with $\xi<0$, $\xi = 0$ and $\xi>0$ corresponding to bounded, exponential and heavy tails, respectively. In practice, for an observed random variable with a finite sample size, a high threshold $u$ is selected and a GPD is fitted to the positive exceedances: we write $X - u \mid X>u \sim \text{GPD}(\tau,\xi)$. 

In many contexts, such as financial and environmental modelling, datasets exhibit non-stationarity, whereby the underlying distribution changes with time or other covariates. In most such cases, we can no longer expect a stationary GPD model to capture the tail adequately. This feature can be present in a range of different forms, as demonstrated by the seasonal and long term trends present in the UKCP18 dataset introduced in Section \ref{sec:intro}. \citet{Davison1990} addressed this issue by using covariates to capture trends in the GPD parameters. Given a non-stationary process $\{Y_t\}$ with covariates $\mathbf{Z}_t$, a non-stationary GPD model is given by  
\begin{equation} \label{eqn:NS_gpd}
    (Y_t - u \mid Y_t > u, \; \mathbf{Z}_t = \mathbf{z}_t) \sim \text{GP}(\tau(\mathbf{z}_t),\xi(\mathbf{z}_t)),  
\end{equation}
for a sufficiently large threshold $u$, where $\tau(\mathbf{z}_t),\xi(\mathbf{z}_t)$ are specified via link functions and linear predictors in the covariates. More recent extensions to this model also allow the threshold $u$ to be covariate dependent. For example, \citet{Kysely2010} and \citet{Northrop2011} use quantile regression to estimate a threshold with a constant exceedance probability, whereas \citet{Sigauke2017} use a cubic smoothing spline. More flexible approaches have been proposed using generalised additive models (GAMs) to capture non-stationary behaviour in univariate extremes \citep{Chavez-Demoulin2005,Youngman2019}. GAMs use smooth functions to capture trends due to covariates, and are less rigid than standard regression models. The general GAM formulation for a parameter $\zeta(\mathbf{z}_t)$ is given by 
\begin{equation} \label{eqn:general_gam_form}
    h(\zeta(\mathbf{z}_t)) = \beta_0 + \sum\limits_{\kappa=1}^{g}\sum\limits_{p=1}^{P_{\kappa}} \beta_{\kappa p}b_{\kappa p} (z_{\kappa,t}),
\end{equation}
with $h(x)$ denoting a link function and $\beta_0,\beta_{\kappa p} \in \RR$ and $b_{\kappa p}$ denoting coefficients and known basis functions, respectively, for $p \in \{1, \hdots, P_{\kappa}\}, \kappa \in \{1, \hdots, g\}$. For each $\kappa$, $P_{\kappa}$ denotes the basis dimension, corresponding to the flexibility of the spline model. The link function $h$ ensures the correct support for the response variable; taking $\zeta(\mathbf{z}_t) = \tau(\mathbf{z}_t)$, for example, one could set $h(x)=\log(x)$ so that $\tau(\mathbf{z}_t) > 0$. In practice, for continuous covariates, smooth basis functions, or splines, are used with equation \eqref{eqn:general_gam_form} to capture the covariate relationships. Estimation of the spline coefficients is carried via a penalised log-likelihood approach, where roughness penalties are imposed to avoid over-fitting; see \citet{Wood2017} for a detailed overview. A wide range of statistical software is available for fitting GAMs, both in the non-extreme \citep{Wood2021} and extreme \citep{Youngman2022} settings.


All the approaches discussed thus far can only be used to model non-stationarity in the extremes of a process. For many statistics corresponding to joint tail behaviour, such as return curves, one must also be able to capture non-stationary within the body of the data simultaneously. This is because extremes of one variable may occur with average values of another variable; see, for instance, the combined projections data for the 1981-1990 period in Figure \ref{fig:UKCP18}. To address such challenges, a range of pre-processing techniques have been proposed that allow marginal non-stationarity to be captured in the body and tail of a dataset simultaneously \citep{Nogaj2007,Eastoe2009,Mentaschi2016}. For these approaches, covariate functions are used to capture and effectively `remove' non-stationarity from the body of the data. Once removed, any remaining trends in the tail can be captured using any of the methods introduced above. The general set-up of these models is to assume
\begin{equation} \label{eqn:ET09}
    (Y_t \mid \mathbf{Z}_t = \mathbf{z}_t) = \mu(\mathbf{z}_t) + \sigma(\mathbf{z}_t)R_t,
\end{equation}
with $\mu$ and $\log(\sigma)$ as linear functions of covariates. Here, the residual process $\{R_t\}$ is assumed to be approximately stationary, and assigning a distribution to this yields a likelihood for all parameters; \citet{Eastoe2009}, for example, adopt a standard normal distribution, with the option to also include a shape transformation. Covariate functions are selected through an analysis of non-stationary trends within the body. 

Alternative approaches for capturing non-stationary behaviour in the body and tail simultaneously include \citet{Opitz2018}, \citet{Krock2022b} and \citet{Carrer2022}. However, such approaches can often result in either less flexible formulations or stronger modelling assumptions compared to pre-processing techniques. We therefore prefer to adopt the latter techniques when modelling marginal non-stationarity.

\subsection{Bivariate extreme value theory} \label{subsec:mvevt}
We briefly recall approaches to modelling extremes in the stationary bivariate setting. To begin, let $(X,Y)$ be a random vector with respective marginal distribution functions $F_X, F_Y$. Consider the conditional probability $\chi(u) = \Pr(F_Y(Y) > u \mid F_X(X) > u)$ and  define the coefficient $\chi:=\lim_{u \to 1}\chi(u) \in [0,1]$ where this limit exists. The cases $\chi = 0$ and $\chi > 0$ correspond to the aforementioned asymptotic independence and asymptotic dependence schemes, respectively. This distinction is important since many models are suitable for data exhibiting one scheme only.

For mathematical simplicity in the description of extremal dependence, it is common to consider random vectors with standardised marginal distributions. This is achieved in practice through marginal estimation and application of the probability integral transform. 

Classical modelling approaches are based on the framework of multivariate regular variation, and are applicable only to asymptotically dependent data. Given a random vector $(X,Y)$ with standard Fr\'echet margins, we define the radial and angular components to be $V:= X+Y$ and $W := X/V$, respectively. We say that $(X,Y)$ is multivariate regularly varying if, for all Borel subsets $B \in [0,1]$, we have 
\begin{equation*}
    \lim_{v \to \infty} \Pr (W \in B, V > sv \mid V>v) = H(B) s^{-1},
\end{equation*}
for any $s>1$, where $H$ is termed the spectral measure \citep{Resnick1987}, and $H(\partial B) = 0$. This assumption implies that, for large radial values, $V$ and $W$ are independent. The spectral measure captures the extremal dependence structure of $(X,Y)$. It must satisfy the moment constraint $\int_0^1wH(\mathrm{d}w) = 1/2$, but has no closed parametric form. All asymptotically independent distributions have a spectral measure placing mass at the endpoints $\{0\}$ and $\{1\}$ of the unit interval, which is why this modelling framework does not form a useful basis for inference under this scheme \citep{Coles1999}. Moreover, it has been shown that assuming the incorrect form of extremal dependence will lead to unsatisfactory extrapolation in the joint tail \citep{Ledford1997,Heffernan2004}. This has consequently led to the development of flexible modelling approaches that are able to theoretically capture both extremal dependence regimes. 

The first such idea was proposed in \citet{Ledford1996,Ledford1997}. It is assumed that the joint tail of a random vector $(X,Y)$ with standard exponential margins is given by 
\begin{equation} \label{eqn:led_tawn}
 \Pr (X > u, Y > u) = \Pr(\min(X,Y)>u) = L(e^u)e^{-u/\eta} \hspace{.5em} \text{as} \; u \to \infty,   
\end{equation}
where $L$ is a slowly varying function at infinity, i.e., $\lim_{x \to \infty}L(cx)/L(x) = 1$ for $c>0$, and $\eta \in (0,1]$. The parameter $\eta$ is termed the coefficient of tail dependence, with $\eta=1$ and $\lim_{u \to \infty}L(e^u) > 0$ corresponding to asymptotic dependence and $\eta < 1$, or $\eta = 1$ and $\lim_{u\to \infty}L(e^u) = 0$, corresponding to asymptotic independence. In Figure \ref{fig:UKCP18}, our estimates of $\eta$ suggest asymptotic independence is exhibited by the UKCP18 data throughout most of the observation period. In practice, this framework is limited by the fact it only characterises the joint tail where both variables are large, and hence is not applicable in regions where only one variable is extreme.

Alternative characterisations of the joint tail have been proposed to circumvent this issue. \citet{Heffernan2004} introduce a general, regression-based modelling tool for conditional probabilities. Given a random vector $(X,Y)$ with standard Laplace margins \citep{Keef2013}, it is assumed that normalising functions $a:\RR \to \RR$ and $b:\RR \to (0,\infty)$ exist such that the following convergence holds: 
$$\lim_{u \to \infty}\Pr\left[(Y-a(X))/b(X) \leq z, X-u > x \mid X>u\right] = D(z)e^{-x}, \hspace{.5em} x>0,$$ 
for a non-degenerate distribution function $D$. Both regimes can be captured via the functions $a$ and $b$, with asymptotic dependence arising when $a(x) = x$ and $b(x) = 1$. Note that one could instead condition on the event $Y>u$. The functions $a$ and $b$ are typically estimated parametrically, while the distribution function $D$ is estimated non-parametrically. This model has been widely used in practice, with applications ranging from air pollution monitoring \citep{Heffernan2004} to coastal flood mitigation \citep{Gouldby2017}.

\citet{Wadsworth2013} provide an alternative representation for the joint tail using a general extension of the framework described in equation \eqref{eqn:led_tawn}. Given $(X,Y)$ with standard exponential margins, they assume that for each $w \in [0,1]$,
\begin{equation} \label{eqn:wads_tawn}
    \Pr(\min\{X/w,Y/(1-w)\}>u) = L(e^u;w)e^{-\lambda(w)u},  \; \; \lambda(w) \geq \max(w,1-w),
\end{equation}
as $u \to \infty$, where $L(\cdot \; ;w)$ is slowly varying for each ray $w \in [0,1]$ and $\lambda$ is the termed the angular dependence function (ADF). This function, which describes the dependence structure of the joint tail along the ray $w$, generalises the coefficient $\eta$, with $\eta = 1/\{2\lambda(0.5)\}$. Both extremal dependence regimes can be captured by $\lambda$, with asymptotic dependence implying the lower bound $\lambda(w) = \max(w,1-w)$ for all $w \in [0,1]$. Pointwise estimates of the ADF can be obtained in practice via the Hill estimator \citep{Hill1975}. Moreover, $\lambda$ captures the joint tail behaviour of a wide range of data structures \citep{Wadsworth2013}, and thus equation \eqref{eqn:wads_tawn} provides a flexible modelling framework for bivariate extremes.

Smooth estimation of the ADF in the stationary setting was recently considered by \citet{Murphy-Barltrop2023a}. Here, the authors show that smooth, functional estimations of $\lambda$ outperform pointwise approaches, such as the Hill estimator, in terms of variance and mean squared error. Furthermore, the authors also derive a shape constraint for the ADF which is outlined in Property \ref{prop:kappa_set}.

\begin{property} \label{prop:kappa_set}
    For any $w_1,w_2 \in [0,1]$ such that $w_1 \leq w_2$, we have 
    \begin{equation*}
        w_1/\lambda(w_1) \leq w_2/\lambda(w_2) \hspace{1em} \text{and} \hspace{1em} (1-w_1)/\lambda(w_1) \geq (1-w_2)/\lambda(w_2). 
    \end{equation*}
\end{property}


Proof of Property \ref{prop:kappa_set} can be found in \citet{Murphy-Barltrop2023a}. In our approach, we exploit this result to ensure the estimated ADFs are theoretically viable; see Section \ref{subsec:theoretical_results} for further details.

Alongside these approaches, we note that there exist several copula-based models that can theoretically capture both extremal dependence regimes, such as those given in \citet{Coles2002}, \citet{Wadsworth2017} and \citet{Huser2019}. However, due to the stronger assumptions about the form of parametric family for the bivariate distribution, we prefer instead to use more flexible modelling techniques. 

\subsection{Non-stationary extremal dependence} \label{subsec:ns_extreme_dep}
Although many extreme value analyses seek to capture marginal non-stationarity, common practice is to assume stationarity in dependence, often without even assessing this feature. Relatively little consideration has been given to this problem in the literature, and most of the approaches that do exist rely on the multivariate regular variation framework, thereby being restricted to asymptotically dependent data. For example, \citet{Mhalla2017} and \citet{Mhalla2019} propose semi-parametric models to capture trends in parameters of quantities related to the spectral measure, while \citet{Carvalho2014}, \citet{Castro-Camilo2018} and \citet{Mhalla2019a} propose flexible modelling techniques for capturing non-stationary trends in the spectral measure under covariate influence. 

\citet{Mhalla2019} also propose a technique for data exhibiting asymptotic independence, using GAMs to capture trends in the non-stationary extension to the ADF defined below in equation \eqref{eqn:ns_wads_tawn}. Given a non-stationary process $\{X_t,Y_t\}$ with standard exponential margins, an external $g$-dimensional covariate $\mathbf{Z}_t$ and any ray $w \in (0,1)$, the extended ADF $\lambda(w \mid \mathbf{z}_t)$ is assumed to take the form described in equation \eqref{eqn:general_gam_form}. The link function $h(x) = \log[ \{ x - \max(w,1-w) \}/(1-x) ]$ is used, resulting in fitted values contained in the interval $[\max(w,1-w),1]$; however, this range is restrictive since $\lambda(w)\leq 1$ implies positive extremal association. In practice, \citet{Mhalla2019} only applied their modelling framework along the ray $w = 1/2$, corresponding to modelling non-stationarity in $\eta$ only. Furthermore, this technique is applied pointwise across each ray $w \in (0,1)$, resulting in non-smooth estimators of the ADF.

Non-stationary extensions to the \citet{Heffernan2004} model also exist: \citet{Jonathan2014a} propose smooth covariate functions for $a$ and $b$, while \citet{Guerrero2021} allow these parameters to vary smoothly over time for blocks of observations via a penalised log-likelihood. However, we note that conditional extremes techniques have been shown to create additional complexities during implementation, requiring more steps compared to alternative approaches because of the need to condition on each variable being extreme separately; see \citet{Murphy-Barltrop2023}. Our proposed method is simpler to implement in practice compared to the those derived under this framework. 

\section{Non-stationary angular dependence function} \label{sec:novel_model}

\subsection{Introduction} \label{subsec:novel_model}
We describe a non-stationary extension to the ADF $\lambda$ of \citet{Wadsworth2013}, which is the key building block for estimating non-stationary return curves. We assume stationary marginal distributions throughout this section, allowing us to separate out the two forms of trends; see  Section \ref{sec:case_study} for further discussion on the separate treatment of these trends. 

Let $\{X_t,Y_t\}$ denote a non-stationary process with stationary, standard exponential marginal distributions. If this is not the case in practice, standard exponential margins can be obtained by first fitting non-stationary marginal distributions, such as those described in Section \ref{subsec:unievt}, and then applying the probability integral transform. Given covariates $\mathbf{Z}_t$, we assume that for all $w \in [0,1]$ and $t \in \{1,2,\ldots,n\}$,
\begin{equation}\label{eqn:ns_wads_tawn}
    \Pr\left( \min\left\{\frac{X_t}{w},\frac{Y_t}{1-w} \right\} > u \mid \mathbf{Z}_t = \mathbf{z}_t \right) = L(e^u \mid w, \mathbf{Z}_t = \mathbf{z}_t) e^{-\lambda(w \mid \mathbf{z}_t)u}, \hspace{.2em} \; u \to \infty,
\end{equation}
where $L$ denotes a slowly varying function and $\lambda(\cdot \mid \mathbf{z}_t)$ denotes the non-stationary counterpart of the ADF at time $t$. This amounts to assuming that the joint tail of $(X_t,Y_t)\mid\mathbf{Z}_t$ can be captured by equation \eqref{eqn:wads_tawn} for all $t \in \{1,2,\hdots,n\}$: this seems reasonable, given the flexibility of the framework outlined in \citet{Wadsworth2013}.

Define $K_{w,t} := \min\left\{X_t/w,Y_t/(1-w) \right\}$: we refer to this variable as the min-projection. Given $w \in [0,1]$ and $t \leq n$, equation \eqref{eqn:ns_wads_tawn} implies that, for large $u$,
\begin{equation*}
    \Pr\left(K_{w,t} > v \Big\vert K_{w,t} > u, \mathbf{Z}_t = \mathbf{z}_t \right) \approx \exp\{-(v-u)\lambda(w \mid \mathbf{z}_t)\}, \; \; v > u. 
\end{equation*}


However, unlike its stationary counterpart, the non-stationary ADF cannot easily be estimated via the Hill estimator because we typically do not have repeated observations for a covariate realisation; even with repeated observations, the resulting sample sizes would typically be too small for reliable estimation. Furthermore, although local non-stationary extensions to the Hill estimator exist \citep[e.g.,][]{deHaan2021}, such techniques are based on user-specified rolling windows, and selection of window size is often not straightforward in practice.  

\subsection{Quantile-based estimators} \label{subsec:quantile_based_est}

\subsubsection{Basic formulation}

We outline new estimation procedures for the non-stationary ADF that can vary across $\mathbf{z}_t$. Given $w \in [0,1]$ and two quantiles $q_1,q_2$ close to one with $q_1<q_2<1$, consider the positive sequences $\{u_{w,t}\}_{t \leq n}$ and $\{v_{w,t}\}_{t \leq n}$ given by 
\begin{equation}\label{eqn:u_w,t}
       \Pr\left( K_{w,t} > u_{w,t} \mid \mathbf{Z}_t = \mathbf{z}_t \right) = 1-q_1 , \hspace{1em} \Pr\left( K_{w,t} > v_{w,t} \mid \mathbf{Z}_t = \mathbf{z}_t \right) = 1-q_2 , 
\end{equation}
for all $t \leq n$. Assuming strict monotonicity of the cumulative distribution function for $K_{w,t} \mid (\mathbf{Z}_t = \mathbf{z}_t)$, we deduce that $v_{w,t} - u_{w,t} > 0$ for all $t \leq n$. Furthermore, the quantile $q_1$ being close to one implies values of the sequence $\{u_{w,t}\}_{t \leq n}$ are large in magnitude. Under the model assumptions, we can therefore deduce that 
\begin{equation*}
  \frac{1-q_2}{1-q_1} = \Pr\left(K_{w,t} > v_{w,t} \Big\vert K_{w,t} > u_{w,t}, \; \mathbf{Z}_t = \mathbf{z}_t \right) \approx \exp\{-(v_{w,t} - u_{w,t})\lambda(w \mid \mathbf{z}_t)\}, 
\end{equation*}
which is rearranged to give
\begin{equation}\label{eqn:lam_formula}
    \lambda(w\mid \mathbf{z}_t) \approx -\frac{1}{v_{w,t} - u_{w,t}}\log\left( \frac{1-q_2}{1-q_1}\right) ,
\end{equation}
for all $t \leq n$. Hence, estimates of the sequences $\{u_{w,t}\}_{t \leq n}$, $\{v_{w,t}\}_{t \leq n}$ lead to a point-wise estimator for the non-stationary ADF at a given angle $w \in [0,1]$. We denote this estimator by $\hat{\lambda}(\cdot\mid \mathbf{z}_t)$, and describe improvements to its stability in Section \ref{subsec:average_over_quantiles}. 

One can also observe that under the modelling assumptions, we have
\begin{equation}\label{eqn:wt_exp_assum}
    (K_{w,t} - u_{w,t}) \vert (K_{w,t} > u_{w,t}, \; \mathbf{Z}_t = \mathbf{z}_t) \sim \text{Exp}(\lambda(w\mid \mathbf{z}_t)).
\end{equation}
Therefore, given some parametric model for $\lambda(w\mid\mathbf{z}_t)$, the non-stationary ADF can be estimated pointwise via likelihood techniques. This is the approach taken by \citet{Mhalla2019}, with  a GAM formulation used to represent $\lambda$ and coefficients estimated using a penalised likelihood approach. We note that in this case, the authors assume a constant threshold $u_{w,t} = u_w$ for all $t \leq n$. This approach may not be desirable in practice, given non-stationarity in the dependence structure implies non-stationarity in the distribution, and hence quantiles, of $K_{w,t}$. We will revisit estimation under equation \eqref{eqn:wt_exp_assum} in Section \ref{subsec:gam_estimators}.

\subsubsection{Estimating quantiles of the min-projection}\label{subsec:quant_reg}
The sequences $\{u_{w,t}\}_{t \leq n}$, $\{v_{w,t}\}_{t \leq n}$ correspond to extreme covariate-varying quantiles of the univariate min-projection $K_{w,t} \mid (\mathbf{Z}_t = \mathbf{z}_t)$ for each $w \in [0,1]$. Quantile regression methods therefore provide a natural solution to the problem of their estimation. Such techniques have successfully been applied in a variety of contexts, ranging from ecology \citep{Cade2003} to growth charts \citep{Wei2006}. Here, we describe two possible approaches for estimating quantiles of the min-projection variable. Given a value $q \in (0,1)$, the $q$-th quantile of $K_{w,t} \mid (\mathbf{Z}_t = \mathbf{z}_t) \sim F_{K_{w,t} \mid \mathbf{z}_t}$ is
\begin{equation*}
    Q_{K_{w,t} \mid (\mathbf{Z}_t = \mathbf{z}_t)}(q) = \inf \{ x : F_{K_{w,t} \mid \mathbf{z}_t }(x \mid \mathbf{z}_t) \geq q \}.
\end{equation*}
A straightforward approach is to assume that the conditional quantile function is linear in $\mathbf{z}_t$, implying $Q_{K_{w,t} \mid (\mathbf{Z}_t = \mathbf{z}_t)}(q) = \mathbf{z}_t'\boldsymbol{\pi}$, where $\boldsymbol{\pi} \in \RR^g$ denotes a vector of coefficients. This is a fairly standard approach for quantile regression, and the vector $\boldsymbol{\pi}$ is estimated through a minimisation of a suitable loss function; see \citet{Koenker2017} for further details. We refer to this approach as the standard quantile regression procedure henceforth. Note that since we consider $q_1$ and $q_2$ seperately, this method of quantile estimation does not guarantee the correct ordering of quantiles, i.e., $u_{w,t} < v_{w,t}$.

Considering the extreme nature of the probabilities $q_1, q_2$ in question, standard quantile regression approaches may not be the most suitable in this context. Therefore, we also consider an alternative framework for obtaining quantile estimates motivated by extreme value theory. In particular, we assume that 
\begin{equation} \label{eqn:min_proj_gpd}
    (K_{w,t} - u^*_{w,t} \vert K_{w,t} > u^*_{w,t}, \; \mathbf{Z}_t = \mathbf{z}_t) \sim \text{GP}(\tau(\mathbf{z}_t),\xi(\mathbf{z}_t)),
\end{equation}
where $u^*_{w,t}$ represents some base `threshold' quantile level obtained using standard techniques. This corresponds with the model recently proposed by \citet{Andre2023} for modelling non-stationary dependence. Assuming the formulation of equation \eqref{eqn:general_gam_form} for $\tau(\mathbf{z}_t)$ and a constant shape parameter, i.e., $\xi(\mathbf{z}_t) = \xi$, we fit model \eqref{eqn:min_proj_gpd} using the \texttt{evgam} package in the \verb|R| computing language \citep{Youngman2022}. Quantile estimates for the min-projection are then obtained from the fitted GP model, and we refer to this approach as the extremal quantile regression procedure henceforth.

For our approach, both quantile regression techniques are used for estimating the non-stationary ADF. A comparison of results is given in Section \ref{sec:sim_study}, where we find that on average, both approaches appear to give unbiased estimates of $\lambda$.

\subsubsection{Averaging over quantiles} \label{subsec:average_over_quantiles}
Prior to applying the proposed model, one must first select $q_1$ and $q_2$ for estimating quantiles of $K_{w,t}$. This selection represents a bias-variance trade off, as is often observed in applications of extreme value theory: quantiles that are not sufficiently extreme (close to one) will induce bias in results, while quantiles that are too large will result in highly variable estimates. Moreover, considering only a single pair of quantiles will lead to higher variability in ADF estimates. To address these issues, we consider a range of quantile pairs simultaneously and compute an average estimator over these values. Specifically, let $\{(q_{1,j},q_{2,j}) \mid 1 \leq j \leq m\}$ be quantiles near one, with $q_{1,j} < q_{2,j} < 1$ for $j = 1, \hdots, m$. Applying standard quantile regression techniques, the pair $(q_{1,j},q_{2,j})$ is used to derive an estimator $\Hat{\lambda}_j$, as in equation \eqref{eqn:lam_formula}, for each $j$. Our final estimator is derived to be the average of these: 
\begin{equation} \label{eqn:ave_estimator}
    \Bar{\lambda}_{QR}(w\mid \mathbf{z}_t) := \frac{1}{m}\sum_{j=1}^m\hat{\lambda}_j(w\mid \mathbf{z}_t),
\end{equation}
for all $w \in [0,1]$ and $t \leq n$. We define $\Bar{\lambda}_{QR2}$ analogously to be the aggregated estimator obtained using the extremal quantile regression procedure. In unreported simulations, we found these aggregated estimators to outperform estimators obtained from any individual pair of quantiles. Furthermore, a range of quantile sets were compared for the examples discussed in Section \ref{sec:sim_study}, with the resulting ADF estimates showing very little difference in variability or accuracy. Our choices for $m$ and $\{(q_{1,j},q_{2,j})\}$ are detailed in Section \ref{sec:sim_study}. 

\subsection{Bernstein-B\'ezier polynomial smooth estimators} \label{subsec:bern_poly_est}
One drawback of the average estimators $\Bar{\lambda}_{QR}$ and $\Bar{\lambda}_{QR2}$ proposed in Section \ref{subsec:average_over_quantiles} is that they are pointwise for each ray $w \in [0,1]$. This typically leads to non-smooth estimates of the ADF that one would not expect to observe in practice. We therefore extend this estimator to give smooth functional estimates using a parametric family derived from the set of Bernstein-B\'ezier polynomials. These polynomials have been applied in many approaches to estimate Pickands' dependence function \citep{Guillotte2016,Marcon2016,Marcon2017a}, a quantity related to the spectral measure which bears many similarities to the ADF \citep{Wadsworth2013}, as well as approaches for estimating the ADF in the stationary setting \citep{Murphy-Barltrop2023a}. In many such approaches, the following family of functions has been considered
\begin{equation*}
    \mathcal{B}_k = \left\{ \sum_{i=0}^k \alpha_i \binom{k}{i} w^i(1-w)^{k-i} \; : \; \boldsymbol{\alpha} \in [0,1]^{k+1}, \; w \in [0,1] \right\},
\end{equation*}
for some degree $k \in \NN$. However, for any $f \in \mathcal{B}_k$, we have $f(w) \leq 1$ for all $w \in [0,1]$. As such, this family of polynomials can only approximate ADFs representing non-negative dependence in the extremes. Furthermore, we wish to allow for covariate influence in the dependence structure; this corresponds to covariate influence in the coefficient vector $\boldsymbol{\alpha}$. We therefore propose extending this family of polynomials to the following set:
\begin{equation*}
    \mathcal{B}^*_{k,\mathbf{z}_t} = \left\{ \sum_{i=0}^k \beta_i(\mathbf{z}_t) \binom{k}{i} w^i(1-w)^{k-i} \; : \; \boldsymbol{\beta}(\mathbf{z}_t) \in [0,\infty)^{k+1}, \; w \in [0,1] \right\},
\end{equation*}
where $\beta_i: \RR^g \to [0,\infty)$ denote functions of the covariates. For any $t \leq n$, let $\lambda_{BP}(\cdot \mid \mathbf{z}_t) \in \mathcal{B}^*_{k,\mathbf{z}_t}$ represent a form of the non-stationary ADF given by this family of functions. Our objective is to find an estimator $\Bar{\lambda}$ that minimises the equation
\begin{equation} \label{eqn:abs_value}
    \lvert \lambda(w\mid \mathbf{z}_t) - \lambda_{BP}(w\mid \mathbf{z}_t)  \rvert
\end{equation}
over all rays $w \in [0,1]$ and $\mathbf{z}_t$ for $t \leq n$; this is achieved through estimation of the coefficient functions $\beta_i$. Since $\lambda$ is unobserved in practice, we consider the objective function
\begin{equation} \label{eqn:sum_theoretical}
    S_{QR}(\boldsymbol{\theta}) := \frac{1}{\vert \mathcal{W} \vert n}\sum_{w \in \mathcal{W}} \sum_{t=1}^n \left\lvert \Bar{\lambda}_{QR}(w\mid \mathbf{z}_t) - \lambda_{BP}(w\mid \mathbf{z}_t, \boldsymbol{\theta}) \right\rvert,
\end{equation}
with $\mathcal{W} := \{0, 0.01, 0.02, \hdots,0.99, 1\}$ defining a finite set spanning the interval $[0,1]$ and $\boldsymbol{\theta}$ denoting the parameter vector corresponding to the coefficient functions $\beta_0,\beta_1,\hdots,\beta_k$. We also define $S_{QR2}$ in an analogous manner. The intuition here is that $S_{QR}(\boldsymbol{\theta})$, or $S_{QR2}(\boldsymbol{\theta})$, gives an approximation of the absolute value in \eqref{eqn:abs_value} integrated over $w$ and $t$: it is therefore desirable to find values of $\boldsymbol{\theta}$ which minimise $S_{QR}$ and $S_{QR2}$. 

To estimate $\boldsymbol{\theta}$, we must specify the form of coefficient functions. To start, we impose that $\beta_0(\mathbf{z}_t) = \beta_k(\mathbf{z}_t) = 1$ for all $t \leq n$; any function $f \in \mathcal{B}^*_{k,\mathbf{z}_t}$ satisfying these conditions has the property that $f(0) = f(1) = 1$, corresponding to the theoretical end-points of the ADF: $\lambda(0) = \lambda(1) = 1$. For $i \in \{1,2,\ldots,k-1\}$, we assume that $\beta_i(\mathbf{z}_t) = h( \mathbf{z}_t'\boldsymbol{\theta}_i)$, where $h:\RR \to [0,\infty)$ denotes a link-function and $\boldsymbol{\theta}_i \in \RR^g$ denotes a vector of coefficients for each $i$. The entire parameter vector is therefore $\boldsymbol{\theta}:=\{\boldsymbol{\theta}_1,\boldsymbol{\theta}_2,\ldots,\boldsymbol{\theta}_{k-1}\}$, with estimators defined by
\begin{equation*}
    \Hat{\boldsymbol{\theta}}_{QR} = \argmin_{\boldsymbol{\theta} \in \RR^{g(k-2)}} S_{QR}(\boldsymbol{\theta}), \hspace{2em} \Hat{\boldsymbol{\theta}}_{QR2} = \argmin_{\boldsymbol{\theta} \in \RR^{g(k-2)}} S_{QR2}(\boldsymbol{\theta}).
\end{equation*}
Finally, smooth estimators of $\lambda(\cdot\mid \mathbf{Z}_t = \mathbf{z}_t)$ are given by $\Bar{\lambda}_{BP}(\cdot \mid \mathbf{z}_t) := \lambda_{BP}(\cdot \mid \mathbf{z}_t, \boldsymbol{\theta} = \boldsymbol{\Hat{\theta}}_{QR})$ and $\Bar{\lambda}_{BP2}(\cdot \mid \mathbf{z}_t) := \lambda_{BP}(\cdot \mid \mathbf{z}_t, \boldsymbol{\theta} = \boldsymbol{\Hat{\theta}}_{QR2})$. 

\subsection{GAM-based estimators}\label{subsec:gam_estimators}
Following the approach of \citet{Mhalla2019}, equation \eqref{eqn:wt_exp_assum} can be exploited to estimate the non-stationary ADF using a GAM formulation. In particular, we assume $\lambda(\cdot \mid \mathbf{z}_t)$ follows the general form denoted in equation \eqref{eqn:general_gam_form}. The valid choices of link function depends on the form of dependence present. The original link function proposed by \citet{Mhalla2019}, $h(x) = \log[ \{ x - \max(w,1-w) \}/(1-x) ]$, can only be used in the case of non-negative dependence, as noted in Section \ref{subsec:ns_extreme_dep}. Therefore, for negative dependence, we propose instead using $h_w(x) = \log(x - \max(w,1-w))$; this still ensures the lower bound is satisfied, while allowing for $\lambda(w)>1$. 

To obtain estimates of GAM coefficients, we use a custom modified version of the \texttt{evgam} package in the \verb|R| computing language \citep{Youngman2022}, whereby the likelihood function associated with equation \eqref{eqn:wt_exp_assum} is used for parameter estimation. We chose to use the \texttt{evgam} package instead of the code provided by \citet{Mhalla2019} since the former was significantly faster and gave more accurate estimates of the non-stationary ADF. For defining min-projection exceedances, we consider three different approaches. For the first, we follow \citet{Mhalla2019} and set $u_{w,t} = u_w$ for all $t \leq n$, where $u_w$ is estimated empirically from the min-projection observations. For the second and third approaches, $u_{w,t}$ is estimated instead using standard and extremal quantile regression techniques, respectively, applying the same methods as outlined in Section \ref{subsec:quant_reg}. These latter methods account for the underlying non-stationarity in the min-projection variable. 

We also note that in \citet{Mhalla2019}, the authors only considered a single fixed quantile level $q_1$ for defining min-projection exceedances. As discussed in Section \ref{subsec:average_over_quantiles}, aggregated estimators appear to have superior properties to those obtained for a single quantile value. Therefore, we define aggregated estimators in an analogous manner to equation \eqref{eqn:ave_estimator}; these are denoted by $\Bar{\lambda}_{GAM}^*$ for the fixed threshold level $u_w$, and by $\Bar{\lambda}_{GAM2}^*$ and $\Bar{\lambda}_{GAM3}^*$ for thresholds estimated using standard and extremal quantile regression techniques, respectively.  



\subsection{Incorporating theoretical properties} \label{subsec:theoretical_results}
All estimators introduced so far are not required to satisfy the shape constraints on $\lambda$ introduced in Section \ref{subsec:mvevt}. We therefore apply post-processing to ensure the estimated ADFs are theoretically valid. Without loss of generality, we consider $\Bar{\lambda}_{QR}$ and assume that there exists some $t \leq n$ such that the set $\{ w \in [0,1] \mid \Bar{\lambda}_{QR}(w\mid \mathbf{z}_t) < \max(w,1-w)\}$ is non-empty. To ensure the ADF is bounded from below, and satisfies the endpoint conditions $\lambda(0) = \lambda(1) = 1$, we set 
\begin{equation*}
    \Bar{\lambda}_{QR}^*(w\mid \mathbf{z}_t) = \begin{cases}
        \max \left\{ \Bar{\lambda}_{QR}(w\mid \mathbf{z}_t), \max(w,1-w) \right\} \; & \text{for} \; w \in (0,1), \\
        1 \; & \text{for} \; w \in \{0,1\}.
    \end{cases} 
\end{equation*}

To ensure the conditions outlined in Property 
\ref{prop:kappa_set} are satisfied, consider the angular sets given by $\mathcal{W}^{\leq 0.5} = (w^{\leq 0.5}_1,w^{\leq 0.5}_2,\hdots,w^{\leq 0.5}_{51}) := (0.5,0.49,\hdots,0) \subset \mathcal{W}$ and $\mathcal{W}^{\geq 0.5} = (w^{\geq 0.5}_1,w^{\geq 0.5}_2,\hdots,w^{\geq 0.5}_{51}) := (0.5,0.51,\hdots,1) \subset \mathcal{W}$. We propose the following algorithm.
\begin{algorithm}
    \SetAlgoLined
    \caption{Algorithm for imposing Property \ref{prop:kappa_set}.} \label{algo:kappa_alg}
    
    \For{$i \leftarrow 2$ \KwTo $51$}{
        \If{$w^{\leq 0.5}_{i-1}/\Bar{\lambda}_{QR}^*(w^{\leq 0.5}_{i-1}) < w^{\leq 0.5}_{i}/\Bar{\lambda}_{QR}^*(w^{\leq 0.5}_{i})$}{
            set $\Bar{\lambda}_{QR}^*(w^{\leq 0.5}_{i}) := w^{\leq 0.5}_{i}\Bar{\lambda}_{QR}^*(w^{\leq 0.5}_{i-1})/w^{\leq 0.5}_{i-1}$ \;
        }
        \If{$(1-w^{\geq 0.5}_{i-1})/\Bar{\lambda}_{QR}^*(w^{\geq 0.5}_{i-1}) < (1-w^{\geq 0.5}_{i})/\Bar{\lambda}_{QR}^*(w^{\geq 0.5}_{i})$}{
            set $\Bar{\lambda}_{QR}^*(w^{\geq 0.5}_{i}) := (1-w^{\geq 0.5}_{i})\Bar{\lambda}_{QR}^*(w^{\geq 0.5}_{i-1})/(1-w^{\geq 0.5}_{i-1})$ \;
        }
         
    }
\end{algorithm}

\begin{figure}[tb]
    \centering
    \includegraphics[width=.95\textwidth]{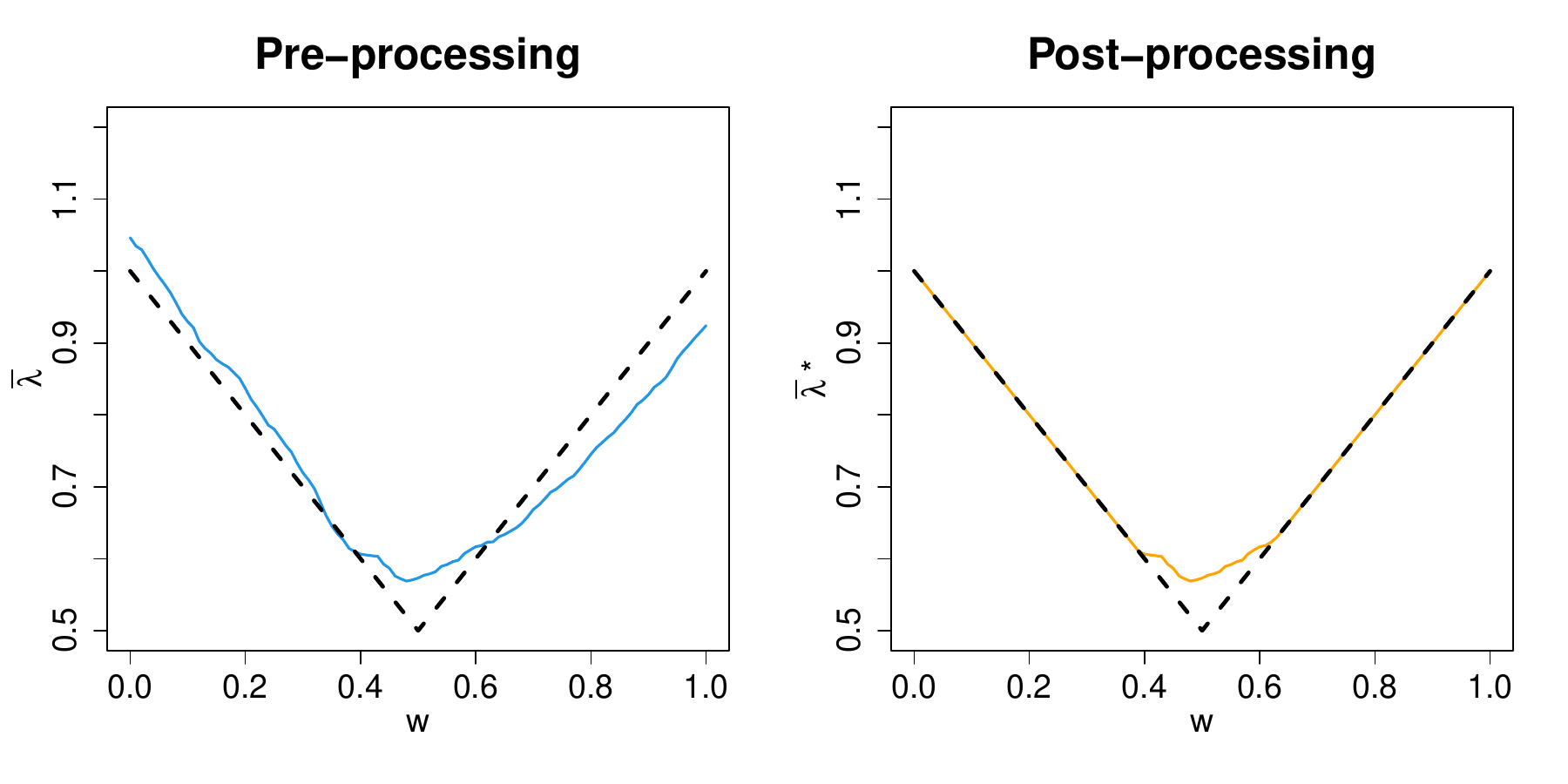}
    \caption{A single ADF estimate before (blue, left) and after (orange, right) processing. The black dotted lines denote the lower bound $\max(w,1-w)$. }
    \label{fig:processing_ADF}
\end{figure}

This ensures the processed estimator $\Bar{\lambda}_{QR}^*$ satisfies the conditions of Property \ref{prop:kappa_set} for all angles in $\mathcal{W}$, which is the finite window that we use to represent the interval $[0,1]$ in practice. We define other processed estimators, denoted $\Bar{\lambda}_{QR2}^*, \Bar{\lambda}_{BP}^*$, $\Bar{\lambda}_{BP2}^*$, $\Bar{\lambda}_{GAM}^*$,  $\Bar{\lambda}_{GAM2}^*$ and  $\Bar{\lambda}_{GAM3}^*$, analogously. In unreported results, we found that imposing these theoretical results improved estimation quality within the resulting ADF estimates. An example ADF estimate before and after processing is illustrated in Figure \ref{fig:processing_ADF}. Observe that imposing Property \ref{prop:kappa_set} via algorithm \ref{algo:kappa_alg} forces the ADF to equal the lower bound for $w \leq 0.39$.  

\subsection{Estimating non-stationary return curves}\label{subsec:NS_RC}

We now consider the problem of estimating $\RC{p}{\mathbf{z}_t}$ at some fixed $t \leq n$ using an estimator of the non-stationary ADF. Let $\lambda^*$ denote one of the defined estimators. Given the set of rays $\mathcal{W}$ defined in Section \ref{subsec:bern_poly_est} and a quantile $q_1$ close to one, we let $\{u_{w,t}\}_{w \in \mathcal{W}}$ be defined as in equation \eqref{eqn:u_w,t}. Then, for all $w \in \mathcal{W}$, define $\{r_{w,t} \}_{w \in \mathcal{W}}$ as 
\begin{equation*}
    r_{w,t} := -\frac{1}{\lambda^*(w\mid \mathbf{z}_t)}\log\left( \frac{p}{1-q_1}\right),
\end{equation*}
implying $p/(1-q_1) = \exp\{-r_{w,t}\lambda^*(w \mid \mathbf{z}_t)\} \approx \exp\{-r_{w,t}\lambda(w \mid \mathbf{z}_t)\}$. Define $(x_{w,t},y_{w,t}) := (w(r_{w,t}+u_{w,t}),(1-w)(r_{w,t}+u_{w,t}))$. We have
\begin{align*}
    \Pr(X_t>x_{w,t},Y_t>y_{w,t} \mid \mathbf{Z}_t = \mathbf{z}_t)  &= \Pr(K_{w,t} > r_{w,t}+u_{w,t} \mid \mathbf{Z}_t = \mathbf{z}_t) \\
    &= \Pr(K_{w,t} > r_{w,t}+u_{w,t} \mid K_{w,t} > u_{w,t}, \mathbf{Z}_t = \mathbf{z}_t)\\
    & \hspace{1em} \times \Pr(K_{w,t} > u_{w,t}\mid \mathbf{Z}_t = \mathbf{z}_t) \\
    &\approx \exp\{-r_{w,t}\lambda^*(w \mid \mathbf{z}_t)\}\Pr(K_{w,t} > u_{w,t}\mid \mathbf{Z}_t = \mathbf{z}_t) \\
    &= \frac{p}{1-q_1}\times 1-q_1 = p,
\end{align*}
meaning that the set $\{(x_{w,t},y_{w,t})\}_{w \in \mathcal{W}}$ provides an approximation of $\RC{p}{\mathbf{z}_t}$. Similarly to the estimation of $\Bar{\lambda}^*_{QR}$ and $\Bar{\lambda}^*_{BP}$, as described in Sections \ref{subsec:average_over_quantiles} and \ref{subsec:bern_poly_est}, we denote $\widehat{\mathrm{RC}}^j_{\mathbf{z}_t}(p) = \{(x^j_{w,t},y^j_{w,t})\}_{w \in \mathcal{W}}$ for each quantile $q_{1,j}$ close to one, and take our final estimator of the return curve to be $\overline{\mathrm{RC}}_{\mathbf{z}_t}(p) = \{(\sum_{j=1}^mx^j_{w,t}/m,\sum_{j=1}^my^j_{w,t}/m)\}_{w \in \mathcal{W}}$.  


\section{Simulation study} \label{sec:sim_study}


\subsection{Overview}
We use simulation to evaluate the properties of the estimators proposed in Section \ref{sec:novel_model}. Section \ref{subsec:sim_examples} introduces a range of examples exhibiting non-stationary extremal dependence with a single covariate. The variation in dependence structures allows us to assess the relative strengths and weaknesses of each estimator. In Section \ref{subsec:inference}, we consider forms for the covariate and link functions, as well as the GAM formulations, introduced in Section \ref{sec:novel_model}. In Section \ref{subsec:sim_study}, we evaluate the bias and variability that arises from each estimator, finding the smooth polynomial and GAM-based estimators to perform best overall. In Section \ref{subsec:sim_RC}, we briefly illustrate how our proposed estimators can be used to derive estimates of non-stationary return curves. Finally, in Section \ref{subsec:twocov_sim_examples}, we introduce an example exhibiting non-stationary extremal dependence with two covariates, and show that a subset of the proposed estimators can still capture the underlying structure. 


\subsection{Non-stationary dependence structures with a single covariate}

\subsubsection{Dependence structures}\label{subsec:sim_examples}

We now introduce several examples exhibiting non-stationary extremal dependence with a single covariate under asymptotic independence. In each case, the non-stationarity is over the time covariate $t \in \{1, 2, \ldots, n\}$, with $n=10,000$ denoting the sample size. The first two examples are obtained using the bivariate normal copula, for which the dependence is characterised by the coefficient $\rho \in [-1,1]$. For the first example, we take $\rho(t) = (t-1)/(n-1)$, so that $\rho(1) = 0$ and $\rho(n) = 1$, i.e., moving from independence to perfect positive dependence. For the second example, we take $\rho(t) = -0.9 + 0.9(t-1)/(n-1)$, giving $\rho(1) = -0.9$ and $\rho(n) = 0$, i.e., moving from a strong negative correlation to independence. 

For the third, fourth and fifth examples, we use the inverted extreme value copula \citep{Ledford1997} with logistic, asymmetric logistic and H\"usler-Reiss families, respectively. For the logistic and asymmetric logistic, the dependence is characterised by the parameter $r \in (0,1)$, with the degree of positive dependence increasing at $r$ approaches $0$. We take $r(t) = 0.01 + 0.98(t-1)/(n-1)$, hence moving from strong positive dependence at $t=1$ to close to independence at $t=n$. The asymmetric logistic distribution also requires two asymmetry parameters $(\kappa_1,\kappa_2) \in [0,1]^2$ \citep{Tawn1988}: we fix $(\kappa_1,\kappa_2) = (0.3,0.7)$, noting this does not change the overall trend in dependence. The H\"usler-Reiss family is characterised by the dependence parameter $s>0$, with independence and complete dependence obtained as $s$ approaches $0$ and $\infty$, respectively. We take $s(t) = 0.01 + 9.99(t-1)/(n-1)$, resulting in increasing dependence over time. 

\begin{figure}[tb]
    \centering
    \includegraphics[width=.95\textwidth]{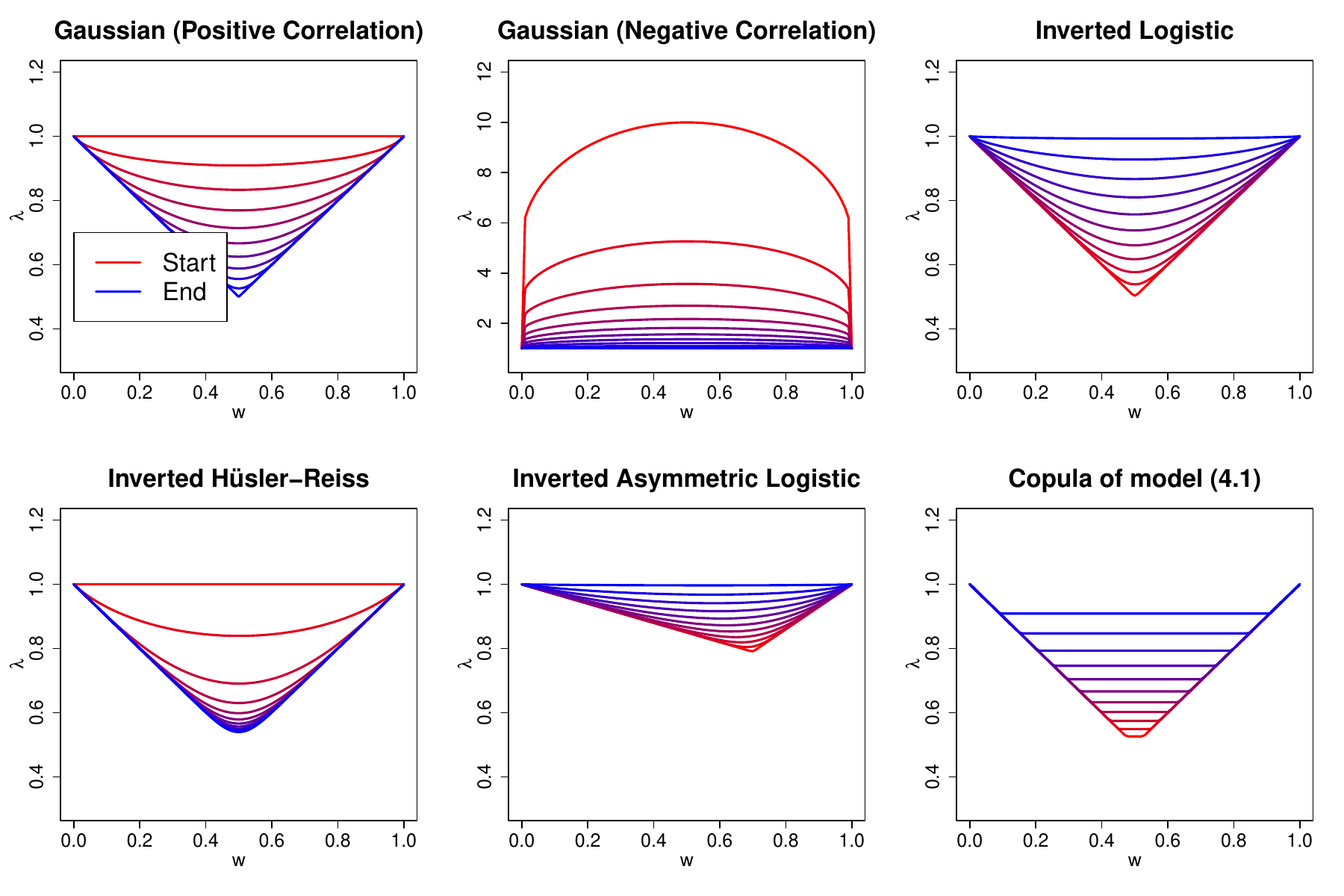}
    \caption{Illustration of true ADFs over time for each copula example. Colour change is used to illustrate trends in extremal dependence structure over the time frame, with red and blue corresponding to the start and end of time frame, respectively.}
    \label{fig:true_ADFs}
\end{figure}

For the final example, we start with a specified ADF and use a method given in \citet{Nolde2022} to construct a copula with this ADF. Given the dependence parameter $c \in (0,1)$, we take $\lambda(w) = \max\{(2w-1)/c,(1-2w)/c,1/(2-c)\}$, which is the ADF of the density proportional to 
\begin{equation}\label{eqn:gauge_dens}
\exp(- \max\{(x-y)/c,(y-x)/c,(x+y)/(2-c)\}),
\end{equation}
and simulate from this density using MCMC. Such a distribution does not have exactly exponential margins, however in this case the transformation to exponential margins, via the probability integral transform, yields a density with the same ADF. We refer to this example as the copula of model \eqref{eqn:gauge_dens} henceforth and set $c(t) = 0.1 + 0.8(t-1)/(n-1)$; this results in a similar dependence trend to the inverted logistic example. Illustrations of the resulting ADFs over time for each example are given in Figure \ref{fig:true_ADFs}. 


\subsubsection{Specification of covariate and link functions, and GAM formulations} \label{subsec:inference}
All estimators proposed in Section \ref{sec:novel_model} require estimates of quantile sequences obtained, which we estimate via one of the proposed quantile regression procedures discussed in Section \ref{subsec:quant_reg}. For the standard quantile regression procedure, one must first specify the functional form for the relationship between quantiles and covariates $\mathbf{z}_t$. Since the data we have simulated has a dependence structure directly related to the covariate $t$, we propose the covariate set $\mathbf{z}_t := \{1,t,t^2,t^3\}$. For any given quantile $q$ and ray $w \in [0,1]$, it is assumed that the quantile function is given by $Q_{K_{w,t} \mid \mathbf{Z}_t = \mathbf{z}_t}(q) = \mathbf{z}_t'\boldsymbol{\pi}$, where $\boldsymbol{\pi} \in \RR^4$. Additional polynomial terms were considered, but we found that a cubic expression was flexible enough to accurately capture quantiles trends for each of the studied examples. Applying the methodology described in \citet{Koenker2017}, estimates of the sequences described in equation \eqref{eqn:u_w,t} can be obtained for any pair of quantiles $q_1,q_2$. 

For the extremal quantile regression procedure, the base quantile level $u_{w,t}$ is estimated using the aforementioned standard technique. The model described in equation \eqref{eqn:min_proj_gpd} is then fitted to the positive exceedances using \texttt{evgam}. For the smooth formulation, we propose using the covariate set $\mathbf{z}_t := \{t\}$ and a cubic regression spline of dimension $15$ for the scale parameter $\tau$. We found that alternative formulations produced little difference in the resulting scale parameters estimates for the examples introduced in Section \ref{subsec:sim_examples}. Quantile estimates of the form denoted in equation \eqref{eqn:u_w,t} are then obtained from the fitted model. 

As noted in Section \ref{subsec:average_over_quantiles}, the choice of quantile sets did not appear to significantly alter the resulting ADF estimates, so long as a range of different quantile pairs were considered. Therefore, for the quantile regression and Bernstein polynomial based estimators, we take $\{q_{1,j}\}^m_{j = 1}$ to be $m=30$ equally spaced points in the interval $[0.9,0.95]$, and set $q_{2,j} = q_{1,j} + 0.04$ for $j = 1, \ldots, m$. We also set the base non-exceedance probability level required for the model described in equation \eqref{eqn:min_proj_gpd} to be $q^* = 0.9$. Moreover, for the GAM based estimators, we similarly take $\{q_{1,j}\}^m_{j = 1}$ to be $m=10$ equally spaced points in the interval $[0.9,0.95]$. Fewer quantiles were considered in this case due to the higher computational intensity of the \texttt{evgam} software, alongside the fact the estimated ADF functions were less variable across different quantile pairs and time points compared to the quantile-based estimates. 


For $\Bar{\lambda}_{QR}^*$ and $\Bar{\lambda}_{QR2}^*$, estimates of non-stationary ADFs can be derived directly using estimated sequences, while specification of coefficient functions are also required for $\Bar{\lambda}_{BP}^*$ and $\Bar{\lambda}_{BP2}^*$. Defining $\mathbf{z}_t := \{1,t\}$, we set $\log(\beta_i(\mathbf{z}_t)) = \mathbf{z}_t'\boldsymbol{\psi}_i$, with $\boldsymbol{\psi}_i \in \RR^2$ for each $i \in \{1,2,\hdots,k-1\}$, thereby ensuring positive coefficient functions. We found this form to be flexible enough to capture the range of dependence trends described in Section \ref{subsec:sim_examples}. 

Finally, for the GAM-based estimators proposed in Section \ref{subsec:gam_estimators}, the link function $h(x) = \log[ \{ x - \max(w,1-w) \}/(1-x) ]$ is used for all non-negative dependence structures, while for the Gaussian example with negative dependence, we use the adapted link function $h_w(x) = \log(x - \max(w,1-w))$. We observe that imposing the upper bound $\lambda(w)=1$ for the GAM-based estimators in the majority of considered examples results in these estimators having a practical advantage over the remaining estimators, which can exceed this upper bound. Furthermore, we set $\mathbf{z}_t := \{t\}$ and specify a thin plate regression spline of dimension $10$ for $\lambda$; this appeared to be sufficient for capturing the range of dependence trends. 

We remark that while the proposed formulations in this section are sufficient for copula examples with a single covariate, adaptations may be required to capture the more complex data structures; see Section \ref{subsec:twocov_sim_examples} for further discussion. 

\subsubsection{Results for one covariate examples} \label{subsec:sim_study}
We now apply each of the estimators introduced in Section \ref{sec:novel_model} to the copula examples discussed in Section \ref{subsec:sim_examples}. To evaluate the performance of the estimators, the mean integrated squared error (MISE), alongside the integrated squared error (ISE) of a median estimator, was computed for each simulated example at three fixed time points; $t=1$, $t=n/2$ and $t=n$. These results can be found in the Supplementary Material. These metrics indicate that, on average, the estimators obtained via the extremal quantile regression procedure outperform their standard quantile regression counterparts, with the Bernstein polynomial and GAM-based estimators performing best overall. Furthermore, while the ISE values for the median estimators from Section \ref{subsec:quantile_based_est} are often competitive in terms of bias, these approaches rarely exhibit the lowest MISE due to their larger variability.

\begin{figure}[tb]
    \centering
    \includegraphics[width=.95\textwidth]{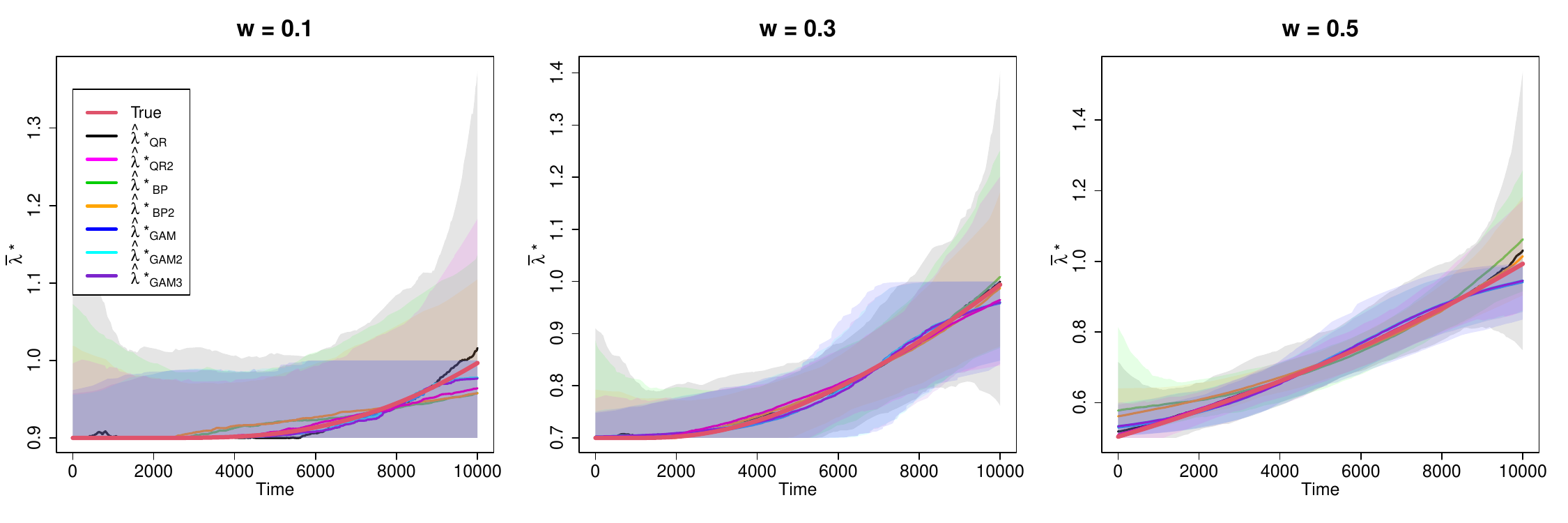}
    \caption{Plots of median and 95\% confidence interval estimates over time at rays $w \in \{0.1,0.3,0.5\}$ for the inverted logistic copula.}
    \label{fig:invlog_bias}
\end{figure}

To further assess the properties of the estimators, we consider the time frame as a whole and we fix rays $w \in \{0.1,0.3,0.5\}$. Using the 250 simulated examples, median estimates of the ADF, alongside $0.025$ and $0.975$ quantiles, are calculated for each ray and time point $t \in \{1,\hdots,n\}$ and these estimates are then plotted over time for fixed rays. The resulting plots for the inverted logistic copula are illustrated in Figures \ref{fig:invlog_bias}. As can be observed, the median estimates for each estimator appear close to the true ADF values, and the bias of each estimator appears similar on average. We note that the uncertainty is noticeably higher for $\Bar{\lambda}^*_{QR}$ and $\Bar{\lambda}^*_{QR2}$, owing to the reduced structure of these estimators. Similar results are obtained for each of the copula examples, with the exception of the negatively correlated Gaussian copula: in this case, significant bias arises nearer the start of the time interval for each estimator, owing to the infinite upper bound for $\lambda$ as $\rho \to -1$. The resulting plots can be found in the Supplementary Material. These results suggest similar bias for each of the estimators proposed in Section \ref{sec:novel_model}. 

Finally, we again fix the time points $t=1, t=n/2$ and $t=n$ and evaluate the variability in ADF estimates. Median curve estimates, alongside pointwise $95\%$ confidence intervals, are obtained for each of the copula examples. As with the analysis of MISE and ISE estimates, these results indicate similar performances across the proposed estimators. Plots of the estimated median curves and confidence regions for each estimator can be found in the Supplementary Material. 

\begin{figure}[H]
    \centering
    \includegraphics[width=\textwidth]{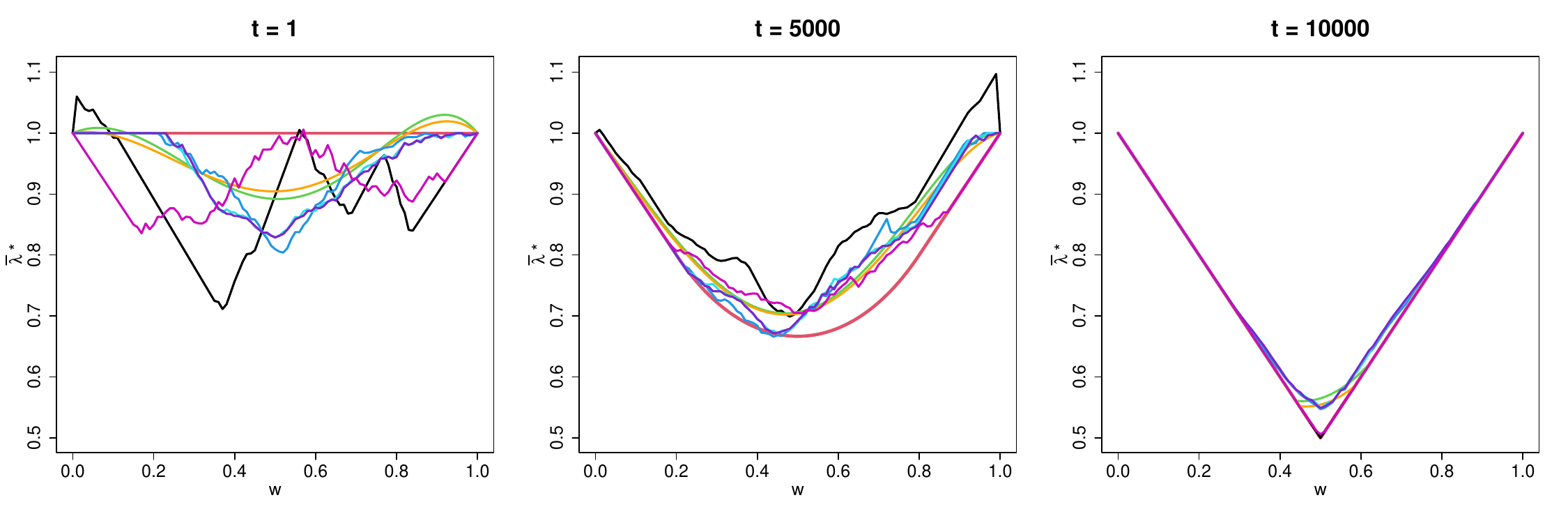}
    \caption{Non-stationary ADF estimates at three fixed time points for a single sample from the Gaussian copula with positive correlation. Legend is as in Figure \ref{fig:invlog_bias}.}
    \label{fig:single_sample_normal}
\end{figure}

While these results indicate a similar performance between the Bernstein polynomial and GAM-based estimators, we believe that the former estimators are preferable in a practical setting. This is due to the fact the GAM-based estimators are pointwise, resulting in non-smooth and unrealistic ADF estimates. To demonstrate this point, Figure \ref{fig:single_sample_normal} illustrates ADF estimates at three time points for all of the proposed estimators; these are computed using single sample from the first copula example discussed in Section \ref{subsec:sim_examples}. One can observe that the pointwise estimators are quite rough, and therefore less realistic. From this, it is apparent that the functional forms of the Bernstein polynomial-based estimators are likely to be preferable in a practical setting. 

\subsubsection{Return curve estimates} \label{subsec:sim_RC}
\begin{figure}[tb]
    \centering
    \includegraphics[width=\textwidth]{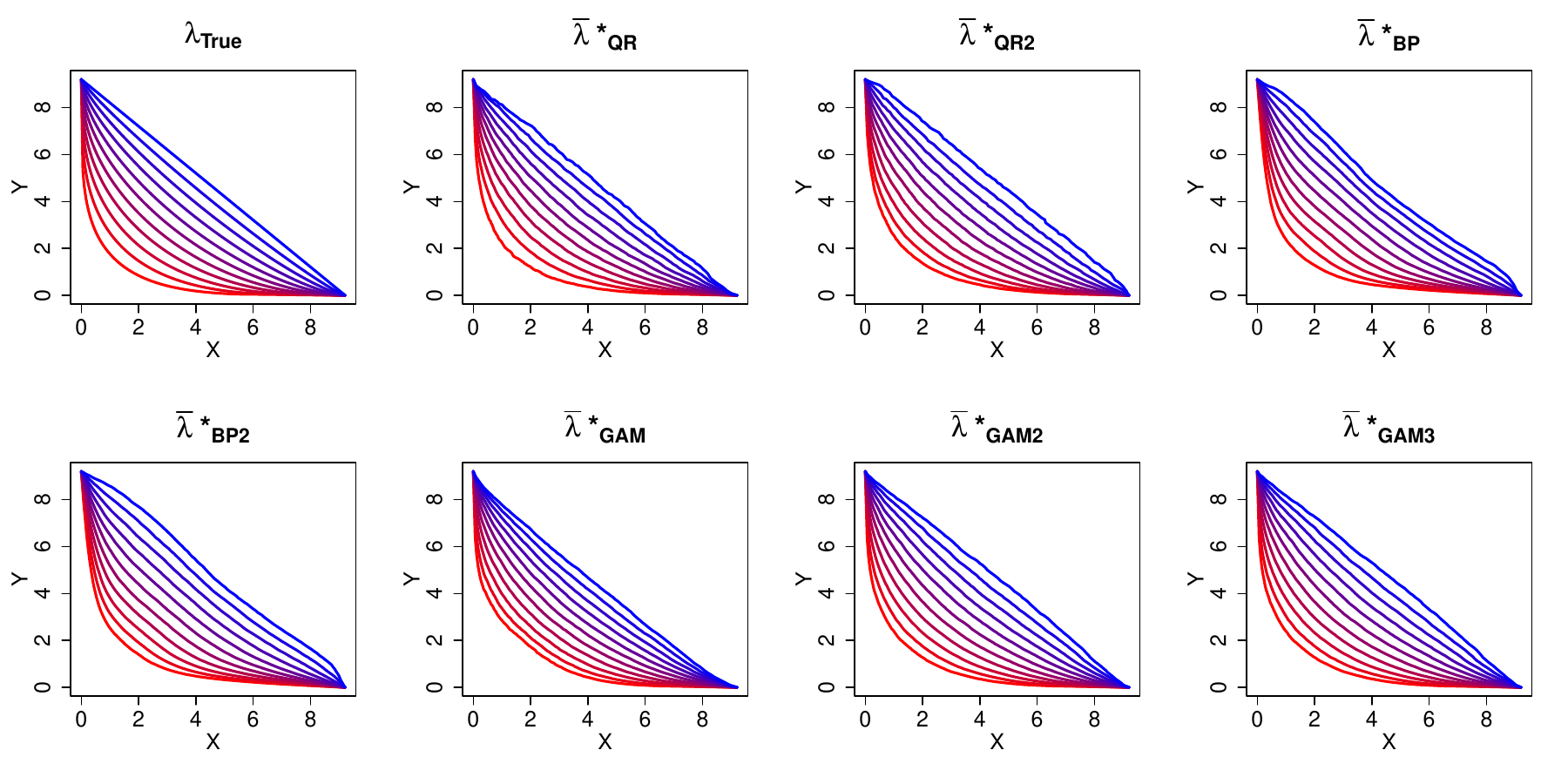}
    \caption{Plots of median return curve estimates over time with $p = 1/n$ for the Gaussian copula with negative correlation. Colour change illustrates extremal dependence trends over time, with red and blue corresponding to start and end of time frame, respectively. True curves given in left panel, with estimated curves from the median estimators of $\Bar{\lambda}^*_{QR}$ and $\Bar{\lambda}_{BP}^*$ given in centre and right panels, respectively.}
    \label{fig:normal_curves}
\end{figure}

We now briefly consider the goal of estimating return curves, $\RC{p}{\mathbf{z}_t}$, for extreme survival probabilities. Curve estimates $\overline{\mathrm{RC}}_{\mathbf{z}_t}(p)$ at $p=1/n$ under a subset of estimators were obtained for 250 simulated examples from each copula. To give an overall impression of the bias from each estimator, we fix a time point $t$ and plot the median of the 250 estimates for $\overline{\mathrm{RC}}_{\mathbf{z}_t}(p)$. Specifically, since each coordinate of $\overline{\mathrm{RC}}_{\mathbf{z}_t}(p)$ is associated with a ray $w \in [0,1]$, we take the median along each ray. Median curve estimates for the negatively correlated Gaussian example are given alongside the true return curves in Figure \ref{fig:normal_curves}. As can be observed, the estimated curves in each case closely resemble the true return curves. This is even true at the start of the observation period, for which significant bias in ADF estimators was observed. Similar plots were obtained for the remaining copula examples.

\subsection{Non-stationary dependence structure with two covariates} \label{subsec:twocov_sim_examples}
To further assess performance, we consider an additional simulated example exhibiting a more complex dependence trend. For this, we only consider a subset of the proposed estimators, specifically $\Bar{\lambda}^*_{BP2}$ and $\Bar{\lambda}^*_{GAM3}$, since these appeared to have the best performance in our one covariate simulation study.


For the simulated example, we define the covariates $a_1(t) := 0.5(t-1)/(n-1)$ and $a_2(t):=\sin(5\pi(t-1)/(n-1))/2$, $t \in \{1, 2, \ldots, n\}$ and take the bivariate normal copula with $\rho(t) = a_1(t) + a_2(t)$, so that $\rho(1) = 0$ and $\rho(n) = 1$. This means that $\rho(t)<0$ for $4611 \leq t \leq 7011$ and $\rho(t)\geq 0$ otherwise, corresponding to a dependence trend that ranges from positive, to negative, then back to positive dependence. A plot of $\rho(t)$ over $t$ can be found in the Supplementary Material. 

Owing to the more complex dependence trends, adaptations to the proposed covariate functions and GAM formulations outlined in Section \ref{subsec:inference} are required. For $\Bar{\lambda}^*_{BP2}$, we take $\mathbf{z}_t := \{1,t,t^2,t^3\}$ and set $\log(\beta_i(\mathbf{z}_t)) = \mathbf{z}_t'\boldsymbol{\psi}_i$, with $\boldsymbol{\psi}_i \in \RR^4$ for each $i \in \{1,2,\hdots,k-1\}$. For $\Bar{\lambda}^*_{GAM3}$, we set $\mathbf{z}_t := \{t\}$ and $h_w(x) = \log(x - \max(w,1-w))$, and specify a cubic regression spline of dimension $15$ for $\lambda$. These adaptations appear to give both estimators sufficient flexibility to capture the structure of the simulated example. 

\begin{figure}[tb]
    \centering
    \includegraphics[width=.95\textwidth]{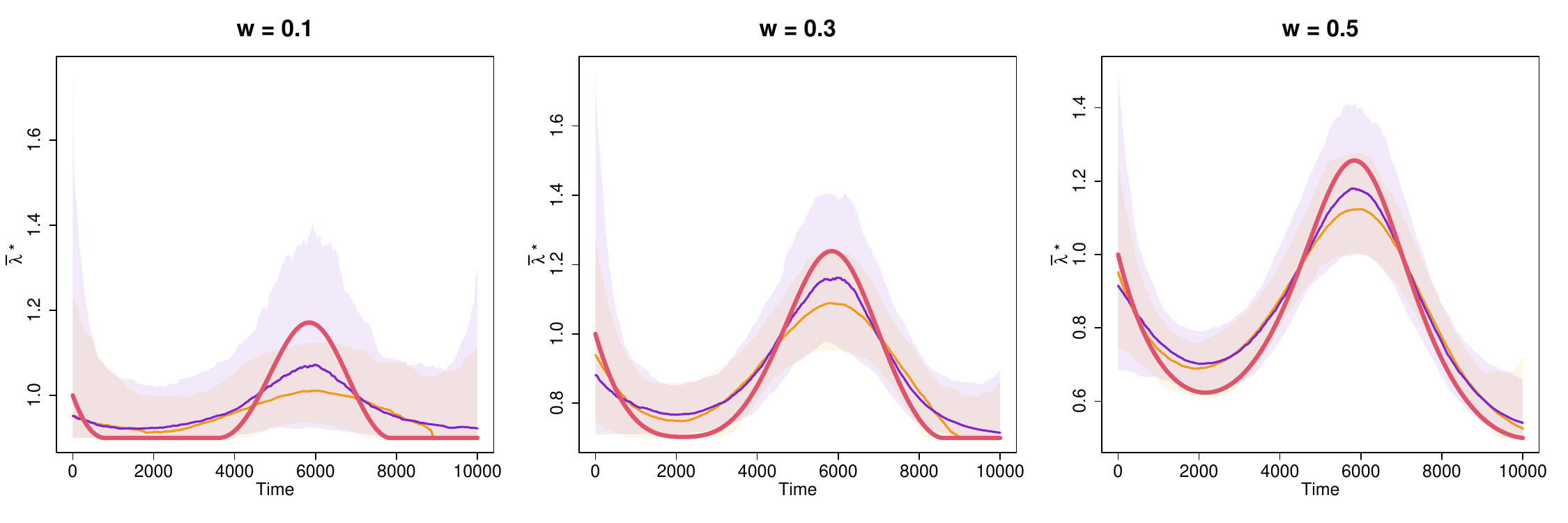}
    \caption{Plots of median and 95\% confidence interval estimates over time at rays $w \in \{0.1,0.3,0.5\}$ for the Gaussian copula with two covariates. Legend as in Figure \ref{fig:invlog_bias}, with the coloured regions representing the estimated confidence intervals.}
    \label{fig:twocov_bias}
\end{figure}

We use the same metrics as outlined in Section \ref{subsec:sim_study} to compare the two estimators. Figure \ref{fig:twocov_bias} gives estimates of the ADF at three fixed angles $w \in \{0.1,0.3,0.5\}$ over time. Tables giving MISE and ISE estimates, alongside plots illustrating ADF estimates for three fixed time points, can be found in the Supplementary Material. From these results, it is clear that the additional flexibility required for the $\Bar{\lambda}^*_{GAM3}$ estimator results in this approach having larger variability then the $\Bar{\lambda}^*_{BP2}$ estimator. 



Overall, these results indicate that $\Bar{\lambda}^*_{BP2}$ is likely to be best suited to capturing more complex dependence structures. Combined with the results discussed in Section \ref{subsec:sim_study}, we recommend this estimator be used in practice when modelling non-stationary dependence, and we restrict attention to $\Bar{\lambda}^*_{BP2}$ for the remainder of this article. 

\section{UKCP18 temperature and dryness data} \label{sec:case_study}

\subsection{Properties of data}

We denote the dataset introduced in Section \ref{sec:intro} as $\{X_t,Y_t\}$ for $t \in \{1,\hdots,n\}$, with $X_t$ and $Y_t$ corresponding to the temperature and dryness variables, respectively. In this case, we have $n=9000$, corresponding to 100 years of summer projections from June 1981 to August 2080. We treat the time index $t$ as a covariate for this data; while this does not correspond to any physical process, it can be used to capture the non-stationarity present in the data, which has been more fully explained by physical inputs to the climate model. For the marginal time series, empirical evidence indicates the presence of seasonal and long term trends within the main bodies of both variables. Further exploratory analysis suggests the presence of non-stationary behaviour within dryness extremes and that non-stationarity is present within the extremal dependence structure, as evidenced by the trend in $\eta$ in Figure \ref{fig:UKCP18}. In this section, we attempt to account for all three forms of non-stationary trends and produce return curve estimates up to the end of the observation period. 

\subsection{Capturing marginal non-stationarity} \label{subsec:marginal_NS}
To capture marginal non-stationarity, we extend the pre-processing technique described in equation \eqref{eqn:ET09}, with the goal of removing any marginal trends from the data. Rather than specifying linear parametric forms for the covariate functions, as in \citet{Eastoe2009}, we instead assume the residual process $R_t$ is a sequence of standard normal variables and allow $\mu$ and $\sigma$ to be general smooth functions of covariate vectors. These functions can be estimated using a GAM framework, allowing for flexible modelling of the marginal trends.

The time series in both cases appear to exhibit long term trends in location and scale, along with seasonal behaviour within the former. Therefore, for the location function $\mu$, and scale function $\sigma$, we take $\mathbf{z}_t = \{1,t,d_t\}$ and $\mathbf{z}_t = \{1,t\}$, respectively, where $d_t \in \{1,2,\ldots,90\}$ denotes the day index of the process at time $t$. For example, $d_t$ equals 1, 45, and 90 for June 1st, July 15th and August 30th, respectively. A thin plate regression spline is used for the covariate $t$, while for $d_t$, a cyclic cubic regression spline of dimension $90$ (corresponding to the number of data points in each year) is used. 


The fitting of the location and scale covariate functions is carried out using the \verb|R| package \verb|mgcv| \citep{Wood2021}, with a \verb|gaulss| family. Model optimisation is achieved via restricted maximum likelihood estimation, with smoothness penalties selected automatically using generalised cross validation. Further details for these modelling procedures can be found in \citet{Wood2017}. The resulting fitted trends are in good agreement with empirical estimates from the marginal time series: see the Supplementary Material for the corresponding plots. Moreover, the transformed series appear stationary, with no obvious long term or seasonal trends in either the location or scale. 

With non-stationary trends in both marginal bodies accounted for, residual processes can be obtained through the transformations $R_{X_t}=[X_t-\hat{\mu}_X(\mathbf{z}_t)]/\hat{\sigma}_X(\mathbf{z}_t)$ and $R_{Y_t}=[Y_t-\hat{\mu}_Y(\mathbf{z}_t)]/\hat{\sigma}_Y(\mathbf{z}_t)$, where $(\hat{\mu}_X,\hat{\sigma}_X)$ and $(\hat{\mu}_Y,\hat{\sigma}_Y)$ denote the estimated covariate functions for $X_t$ and $Y_t$, respectively. Assuming an accurate model fit, these processes should be approximately stationary within the body of the data. However, non-stationary trends may remain in the tails since GAM fitting is driven by the body. Following \citet{Eastoe2009}, we fit the non-stationary GPD described in equation \eqref{eqn:NS_gpd} to capture any remaining trends. Significant linear and harmonic trends are shown to exist within the scale parameter of the residual process for dryness, while the shape parameter is assumed to be fixed over time. This assumption is common within the analysis of non-stationary univariate extremes \citep[e.g.,][]{Eastoe2009,Chavez-Demoulin2012}, since the shape parameter is often seen seen as an intrinsic property of a physical process. No significant trends were found for the scale parameter corresponding to the temperature variable. Let $\{(r_{X_t},r_{Y_t}):t=1,\hdots,n\}$ denote a data sample corresponding to the residuals vector $(R_{X_t},R_{Y_t})$. An estimate of the marginal distribution function $F_{R_{Y_t}}$ is given by
\begin{equation} \label{eqn:CDF_trans}
    \hat{F}_{R_{Y_t}}(r \mid \mathbf{Z}_t = \mathbf{z}_t) = \begin{cases} & 1 - (1 - q_Y)\{1 + \hat{\xi}_Y(r-u_Y)/\hat{\tau}_Y(\mathbf{z}_t) \}^{-1/\hat{\xi}_Y} \; \text{for} \; r > u_Y, \\ & \sum_{t=1}^n \mathbbm{1}(r_{Y_t} \leq r)/(n+1) \; \hspace{7.83em}  \text{for} \; r \leq u_Y,  \end{cases}
\end{equation}
where $\mathbbm{1}$ denotes an indicator function, $u_Y$ is the empirical $q_Y$ quantile of $R_{Y_t}$, and $(\hat{\tau}_Y(\mathbf{z}_t),\hat{\xi}_Y)$ are the MLEs of the GPD scale and shape parameters. The stationary GPD, denoted in equation \eqref{eqn:gpd_cdf}, is used to estimate the upper tail of $F_{R_{X_t}}$. 

Finally, the data is transformed to standard exponential margins via the probability integral transform. To assess the outcome of the pre-processing procedure, exponential rate parameters were estimated over time for both marginal processes. The resulting estimates remain approximately constant at one (the target value) throughout the observation period; an illustrative plot can be found in the Supplementary Material.


\subsection{Model fitting}
With the data transformed to standard exponential margins, we apply the methodology proposed in Section \ref{sec:novel_model} for return curve estimation. The extremal dependence trend observed in Figure \ref{fig:UKCP18} appears to suggest the extremes of the process are becoming more positively dependent over the time frame; therefore, one may expect the ADF estimates to tend towards the lower bound as $t \to n$. 

\begin{figure}[tb]
    \centering
    \includegraphics[width=.9\textwidth]{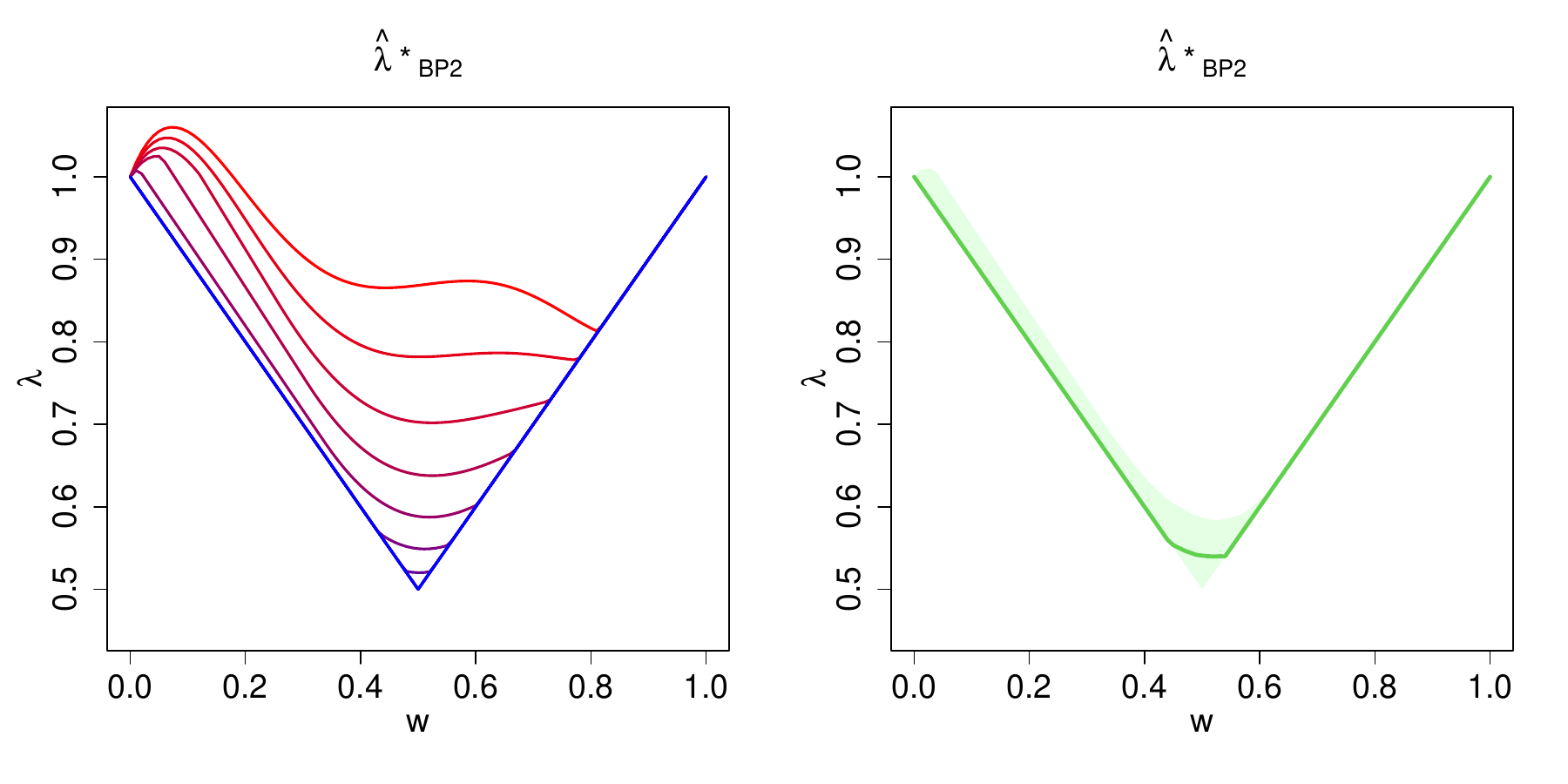}
    \caption{Left: ADF estimates over time for $\Bar{\lambda}_{BP2}^*$. Curves change from red to blue over the observation period. Right: median ADF estimate for $t = n/2$ obtained through bootstrapping procedure, with coloured region representing uncertainty bounds.}
    \label{fig:UKCP18_ADFs}
\end{figure}

To estimate $\Bar{\lambda}_{BP2}^*$, we set $\mathbf{z}_t := \{t\}$ for the extremal quantile regression procedure, and $\mathbf{z}_t := \{1,t\}$ for the estimation of Bernstein polynomial coefficients. These covariate spaces were flexible enough to capture the observed extremal dependence trend within the data. The same set of quantile pairs $\{q_{1,j},q_{2,j}\}_{j=1}^m$ was considered as defined in Section \ref{subsec:inference}. The resulting ADF estimates over the observation period are illustrated in the left panel of Figure \ref{fig:UKCP18_ADFs}. The selected values of $t$ for the plotted curves correspond to July 15th for an increasing subset of equally spaced years between 1981-2080. We remark that the linear behaviour observed for magenta coloured curves on the first half of the angular interval is a direct result of imposing Property \ref{prop:kappa_set} using the algorithm described in Section \ref{subsec:theoretical_results}.

To evaluate uncertainty in estimates, we propose a block bootstrapping procedure. First, the data on exponential margins is split into segments of size $450$, corresponding to five years of observations. The extremal dependence structure within such segments is then assumed to be approximately stationary. Within each segment, data is then resampled in blocks of size $15$ with replacement to account for temporal dependence. These blocks are combined to form a resampled segment, then each of the segments are merged in order to obtain a new dataset. This process is repeated $250$ times to generate sets of ADF estimates while accounting for complex structures in the data. The median of the estimated ADFs for $t=n/2$, alongside $95\%$ pointwise confidence intervals, are illustrated in the right panel of Figure \ref{fig:UKCP18_ADFs}. Note that imposing the shape constraints on the ADF, as described in Section \ref{subsec:theoretical_results}, reduces the range of the confidence intervals, thus explaining why the median estimate appears close to the lower bound. 

Finally, to assess the quality of the ADF estimates, we have compared the model estimates of $\eta$ over time to the empirical estimates introduced in Section \ref{sec:intro}. For each rolling window, we have taken the average $\eta$ estimate from the fitted model. As illustrated in the left panel of Figure \ref{fig:UKCP_ns_curves}, the model estimates appear similar over time, suggesting we have accurately captured the extremal dependence trend at this ray. Similar results were also observed for model estimates at other rays.



\subsection{Return curve estimates} 
\begin{figure}[tb]
    \centering
    \includegraphics[width=.9\textwidth]{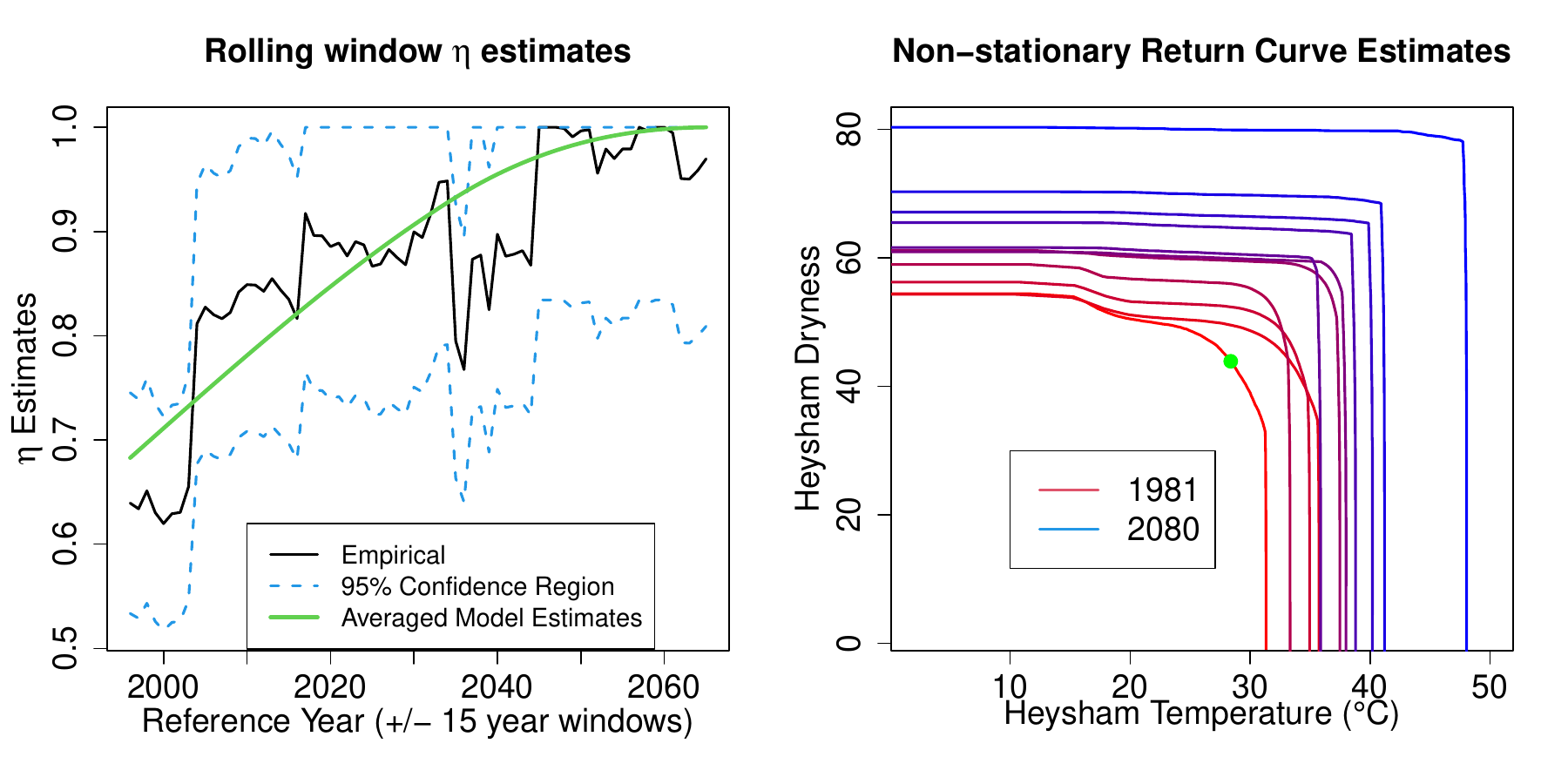}
    \caption{Left: Comparison of averaged model $\eta$ estimates for rolling windows to empirical estimates, with black, green and dotted blue lines corresponding to empirical, model, and 95\% confidence interval estimates respectively. Right: Return curve estimates on original margins with $p=1/n$ at July 15th over the observation period. Time is illustrated using a colour transition, with the curves for the start and end of the time frame labelled. The green point corresponds to the ray $w=0.5$ when the 1981 curve is transformed to standard exponential margins.}
    \label{fig:UKCP_ns_curves}
\end{figure}

We use our estimated ADFs to estimate return curves $\overline{\mathrm{RC}}_{\mathbf{z}_t}(p)$ up to the year 2080. In the stationary setting, we define `1 in 10,000-year' bivariate events to be those with joint survival probability of $p = 1/(10000\times n_y)$, where $n_y$ denotes the average number of observations per year \citep{Brunner2016}. We therefore obtain return curve estimates at this probability level for the vector $(X_t,Y_t)$ using $\Bar{\lambda}_{BP2}^*$. This is done in two steps: first, we apply the method introduced in Section \ref{subsec:NS_RC} to obtain return curve estimates on standard exponential margins. These curves are then transformed back to the original scale by applying the the inverse of the semi-empirical distribution given in equation \eqref{eqn:CDF_trans}, followed by the inverses of the transformations used to obtain the variables $R_{X_t}$ and $R_{Y_t}$. The resulting curve estimates are illustrated in the right panel of Figure \ref{fig:UKCP_ns_curves}, with the selected values of $t$ for the plotted curves again corresponding to July 15th for an increasing subset of years between 1981-2080. All of the theoretical properties for return curves introduced in \citet{Murphy-Barltrop2023} have been imposed to ensure the resulting estimates are both theoretically possible and realistic.

From these curve estimates, two conclusions of relevance to nuclear regulators are evident. Firstly, clear marginal trends can be observed for both time series. For example, the temperature values at the point where the curves intersect the $x$-axis, which equate to $(1-p)$-th non-stationary quantile estimates (i.e., `1 in 10,000-year' univariate events), increase by over $15^\circ$C through the observation period. This implies that the regulatory design values will increase significantly over the time frame. Moreover, a dependence trend is evidenced by the changing shape in return curve estimates over time, with the curves becoming increasingly `square' shaped from 1981-2080. This suggests that joint extremes of temperature and dryness are becoming more likely to occur over the observation period, implying drought-like conditions could be more common at the end of the time frame. 

To better illustrate the shift in joint extremal behaviour over the observation period, consider the point labelled on the 1981 curve in green; this corresponds to the ray $w=0.5$ when translated to standard exponential margins. These coordinates considered in the year 2080 would equate to a one in $0.52$-year joint survival event. Moreover, the marginal return periods in 2080 would be $0.09$ and $0.62$ years for temperature and dryness, respectively. These values are orders of magnitude different from the 10,000-year regulatory standard. At the other end of the scale, values on the 2080 return curve lie above the fitted upper endpoints of the 1981 marginal distributions.


The non-stationarity in these return curves should be taken into account when considering the design basis for future nuclear installations -- in particular, conservative principles suggest designing to values occurring at the end of the time frame. Moreover, due to the observed change in extremal dependence, future designs must be able to cope with the most extreme values of temperature and dryness occurring simultaneously, again resulting in a more complex design specification compared to the start of the observation period. 

We note that there is a large degree of uncertainty in these curve estimates which cannot be quantified in a simple manner - see Section \ref{sec:discussion} for further discussion. Uncertainty arises in every step of the estimation procedure; see, for example, the estimated uncertainty for the fitted GAM functions illustrated in the Supplementary Material. Marginal trends play a key role in the changing values of the return curves, and differences in these point estimates would naturally impact the estimated curves.

\section{Discussion} \label{sec:discussion}
We have proposed a novel method for capturing non-stationary extremal dependence structures under asymptotic independence. Our method has been successfully applied to heavily non-stationary data from the UKCP18, allowing us to obtain return curve estimates up to the year 2080 and thereby illustrating how this framework could help improve practical risk management under future climate scenarios. 

We have investigated the properties of our estimators via simulation and have observed them to generally perform well in terms of bias. While it would be desirable also to have theoretical results on bias of the estimators, this is very challenging in practice and is likely to require assumptions that are too strict to make the results worthwhile. Potentially a more promising line of future work would be the development of diagnostic plots for non-stationary ADF and return curve estimates. Our diagnostic procedures were limited to comparison of rolling window $\eta$ estimates against those derived from the ADF. A further possibility would be the exploit the fact $(K_{w,t} - u_{w,t})\lambda(w \mid \mathbf{z}_t)$, $K_{w,t} > u_{w,t}$, as defined in Section \ref{sec:novel_model}, should follow a standard exponential distribution for each ray $w$. The use of diagnostic plots could also be helpful for informing covariate selection. In general, existing diagnostic plots for return curves, such as those introduced in \citet{Murphy-Barltrop2023}, require stationarity assumptions. 


For the sake of simplicity, we have restricted attention to the bivariate setting. However, it is worth noting that all methods could in principle be extended to the general multivariate setting. This scenario results in additional complexities, since different extremal dependence regimes can exist within subvectors of a multivariate random vector, and more sophisticated model formulations may be required to capture such dependence structures. Moreover, applications of multivariate extremal risk measures are limited, owing in part to the fact visualisation and interpretation becomes more challenging in higher dimensions.




As noted in Section \ref{subsec:theoretical_results}, imposing the theoretical result developed by \citet{Murphy-Barltrop2023a} on the estimates of the ADF was shown to improve the estimation procedure. This was an ad-hoc post-processing step, with many of the obtained ADF estimates not satisfying the shape constraints. Future work could explore how these constraints could be incorporated directly into the modelling framework, and whether this would further improve the quality of ADF estimates. We note that when asymptotic dependence is present, $\lambda(w \mid \mathbf{z}_t) = \max(w,1-w)$ will cease to depend on $\mathbf{z}_t$. However, return curve estimates $\overline{\mathrm{RC}}_{\mathbf{z}_t}(p)$ are also affected by the sequences $\{u_{w,t}\}$ as described in Section \ref{subsec:NS_RC}, so will still capture non-stationarity. Nonetheless, if asymptotic dependence is clearly present at all time points, other techniques for estimating non-stationarity may be preferable.

We recognise that the method proposed for evaluating and representing uncertainty in Section \ref{sec:case_study} relies on strong assumptions and is an approximation of the true uncertainty. Theoretical derivation of uncertainty intervals for either of the estimators proposed in Section \ref{sec:novel_model} is not possible, meaning any evaluation of uncertainty must be non-parametric. Uncertainty quantification is a general problem when modelling non-stationary processes, since the underlying datasets cannot be resampled using a straightforward bootstrapping procedure. 



Finally, we note that the data within the climate projections exhibits non-negligible temporal dependence. This feature decreases the amount of information available, and we found that the ADF estimation procedures detailed in Section \ref{sec:novel_model} performed worse when this feature was present. However, for ease of implementation, we have assumed independence in both marginal distributions. Techniques for incorporating temporal dependence with quantile regression \citep{Koenker2017} could be incorporated into our methodology in future work.  

\section*{Acknowledgements}
We are grateful to the two referees for constructive comments and suggestions that have improved this article. We would also like to thank Ben Youngman for his assistance with the \texttt{evgam} package in the \verb|R| computing language.

\section*{Funding}
This paper is based on work partly completed while Callum Murphy-Barltrop was part of the EPSRC funded STOR-i centre for doctoral training (EP/L015692/1). 

\section*{Conflict of interest}
The authors declare that they have no conflict of interest.

\section*{Data availability}
The dataset analysed during the current study is available from the corresponding author on reasonable request.

\bigskip
\begin{center}
{\large\bf SUPPLEMENTARY MATERIAL}
\end{center}

\textbf{Supplementary Material for ``Modelling non-stationarity in asymptotically independent extremes".} We provide additional figures relevant to the UKCP18 dataset; in particular, plots related to the varying extremal dependence across seasons and modelling of marginal non-stationarity. We also include additional supporting results for the simulation study described in Section \ref{sec:sim_study}, which further illustrate ADF and return curve estimates across the considered examples. (.pdf file) \\
\indent \textbf{Case study program and data availability.} We provide an \textit{R} program to generate the case study plots discussed in Section \ref{sec:case_study}, along with the UKCP18 projections. These plot relate to estimation of the marginal distributions and of non-stationary ADFs and return curves via the constrained version of $\Bar{\lambda}_{BP2}^*$. (.zip file)

\bibliography{LaTeX/library.bib,LaTeX/additional}       

\pagebreak

\begin{center}
\textbf{\large Supplementary Material to `Modelling non-stationarity in asymptotically independent extremes'}
\end{center}
\setcounter{equation}{0}
\setcounter{figure}{0}
\setcounter{table}{0}
\setcounter{page}{1}
\setcounter{section}{0}
\makeatletter
\renewcommand{\theequation}{S\arabic{equation}}
\renewcommand{\thefigure}{S\arabic{figure}}
\renewcommand{\thesection}{S\arabic{section}}
\renewcommand{\bibnumfmt}[1]{[S#1]}
\renewcommand{\citenumfont}[1]{S#1}

\section{Varying extremal dependence structures across seasons for UKCP18 dataset}

\begin{figure}[H]
    \centering
    \includegraphics[width=\textwidth]{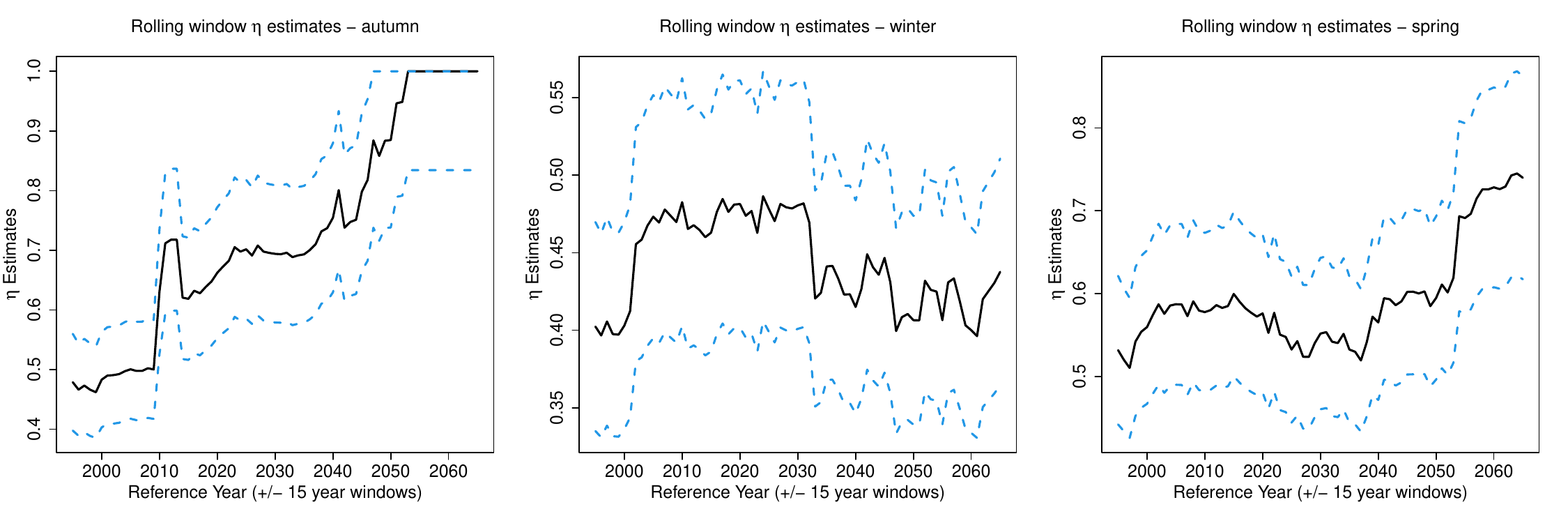}
    \caption{Trends in $\eta$ parameter estimates (solid black lines) over $\pm 15$ year rolling windows for autumn, winter and spring, alongside $95\%$ pointwise confidence intervals (dotted blue lines).}
    \label{fig:eta_seasons}
\end{figure}

Trends in the extremal dependence structure for the UKCP18 projections were considered across each of the meteorological seasons independently. For a given season, the corresponding subset of data was transformed to exponential margins using the same techniques as described in the case study of the main text (Section 5). The coefficient $\eta$ was then estimated across $\pm 15$ year rolling windows over the observation period; the resulting plots for Autumn, Winter and Spring are given in the left, centre and right panels of Figure \ref{fig:eta_seasons}. These plots illustrate significantly different behaviour across these seasons, justifying our choice to just consider summer data within the case study. Moreover, summer is likely to correspond to the highest temperature and dryness values, hence it makes most sense to consider joint extremal behaviour for this season. 

\section{Additional simulation study results for non-stationary dependence structures with a single covariate}

\subsection{MISE values of estimators}
To evaluate the performance of the estimators, estimates of the mean integrated squared error (MISE) were obtained using $250$ samples from each copula. Given an estimator $\Bar{\lambda}^*$, the MISE at time $t$ is given by 
\begin{equation*}
    \text{MISE}(\Bar{\lambda}^*(\cdot \mid \mathbf{z}_t)) = \mathbb{E} \left( \int_0^1 \left[\Bar{\lambda}^*(w \mid \mathbf{z}_t) - \lambda(w \mid \mathbf{z}_t) \right]^2 \mathrm{d}w \right),
\end{equation*}
with smaller MISE values corresponding to estimators with lower bias and variance. Three different time points, $t=1, t=n/2$ and $t=n$, were considered, corresponding to the start, middle and end of the simulated time frame, respectively.

Table \ref{table:ISE_values2} gives MISE values for each estimator and copula example at each time point. One can observe the lowest MISE values are always for Bernstein polynomial or GAM-based estimators; this is likely due to the reduced variance of these approaches compared to quantile-based estimators, owing to the reduced structure of the latter estimators. Moreover, we observe that, on average, the estimators obtained using extremal quantile regression techniques outperform their standard quantile regression counterparts. 

\begin{landscape}
    \renewcommand{\arraystretch}{1.1}
    \begin{table}[ht]
    \caption{MISE values (multiplied by 1,000) at start, middle and end of simulated time frame. Smallest MISE values in each row are highlighted in bold.}
    \label{table:ISE_values2}
    \centering 
        \begin{tabular}[t]{|c|c|c|c|c|c|c|c|c|}
        \hline 
        Copula & Times & $\Bar{\lambda}_{QR}^*$ & $\Bar{\lambda}_{QR2}^*$ & $\Bar{\lambda}_{BP}^*$ & $\Bar{\lambda}_{BP2}^*$ & $\Bar{\lambda}_{GAM}^*$ & $\Bar{\lambda}_{GAM2}^*$ & $\Bar{\lambda}_{GAM3}^*$\\
        \hline
        Gaussian (Positive Correlation) & Start & 25.310 & 6.418 & 10.117 & 4.085 & 4.209 & \textbf{3.326} & 3.426\\
        \hline
        Gaussian (Positive Correlation) & Middle & 2.959 & 2.450 & 2.062 & 2.151 & 2.210 & 2.118 & \textbf{2.033}\\
        \hline
        Gaussian (Positive Correlation) & End & 6.485 & 1.255 & 3.355 & 1.546 & \textbf{0.368} & 0.478 & 0.442\\
        \hline
        Gaussian (Negative Correlation) & Start & 32765.050 & 37324.910 & 35762.060 & 38024.650 & \textbf{31529.260} & 34326.670 & 34545.940\\
        \hline
        Gaussian (Negative Correlation) & Middle & 86.186 & 83.788 & 86.458 & 80.896 & 67.236 & 69.662 & 70.504\\
        \hline
        Gaussian (Negative Correlation) & End & 17.116 & 9.156 & 8.490 & 7.511 & \textbf{3.830} & 4.868 & 4.675\\
        \hline
        Inverted Logistic & Start & 8.513 & 0.576 & 4.826 & 1.010 & \textbf{0.300} & 0.392 & 0.382\\
        \hline
        Inverted Logistic & Middle & 1.485 & 1.000 & 1.041 & \textbf{0.789} & 1.412 & 1.335 & 1.301\\
        \hline
        Inverted Logistic & End & 24.822 & 6.269 & 8.927 & 4.331 & 3.489 & \textbf{2.751} & 2.865\\
        \hline
        Inverted Husler-Reiss & Start & 42.468 & 12.421 & 20.158 & 10.590 & 13.055 & \textbf{5.968} & 6.201\\
        \hline
        Inverted Husler-Reiss & Middle & 0.660 & 0.543 & 0.417 & 0.476 & 0.399 & 0.406 & \textbf{0.392}\\
        \hline
        Inverted Husler-Reiss & End & 8.548 & 2.006 & 3.922 & 1.268 & 0.189 & 0.146 & \textbf{0.145}\\
        \hline
        Inverted Asymmetric Logistic & Start & 16.971 & 4.151 & 6.015 & \textbf{2.922} & 3.050 & 3.206 & 3.273\\
        \hline
        Inverted Asymmetric Logistic & Middle & 2.493 & 1.393 & 1.421 & \textbf{0.897} & 1.475 & 1.547 & 1.527\\
        \hline
        Inverted Asymmetric Logistic & End & 20.761 & 5.574 & 6.408 & 3.578 & 3.605 & \textbf{3.147} & 3.289\\
        \hline
        Copula of model (4.1) & Start & 6.437 & 0.817 & 4.831 & 1.034 & \textbf{0.093} & 0.105 & 0.100\\
        \hline
        Copula of model (4.1) & Middle & 0.964 & 0.765 & 0.810 & \textbf{0.712} & 0.808 & 0.788 & 0.767\\
        \hline
        Copula of model (4.1) & End & 36.102 & 7.744 & 14.856 & 6.434 & 3.300 & 3.140 & \textbf{3.117}\\
        \hline
    \end{tabular}
\end{table}
\end{landscape}

\subsection{ISE values of median estimators}
Using $250$ samples from each copula, median ADF estimators were computed pointwise over the set $\mathcal{W} = \{0, 0.01, 0.02, \hdots, 0.99, 1\}$ for three different time points: $t=1, t=n/2$ and $t=n$, corresponding to the start, middle and end of the simulated time frame, respectively. The integrated squared error (ISE) of the median estimators is used to compare performance across all of the estimators. Although these median estimators are not computable in a given application, understanding their properties gives an insight into the bias of the estimators. Letting $\text{med}_t\Bar{\lambda}^*$ denote a median estimator at time $t$, the ISE is given by 
\begin{equation*}
    \text{ISE}( \text{med}_t\Bar{\lambda}^*(\cdot \mid \mathbf{z}_t)) = \int_0^1 \left[\text{med}_t\Bar{\lambda}^*(w \mid \mathbf{z}_t) - \lambda(w \mid \mathbf{z}_t) \right]^2 \mathrm{d}w,
\end{equation*}
with smaller ISE values corresponding to an estimator with lower bias. Table \ref{table:ISE_values} gives the ISE values for each median estimator and copula example at each of the three time points. As can be observed, the results from each of the estimators are generally similar, and the obtained bias values appear on the same order of magnitude in most cases. Moreover, we note that $\Bar{\lambda}^*_{QR2}$ often results in the lowest ISE values compared to the other estimators, suggesting the extremal quantile regression procedure results in the least bias on average. 


\begin{landscape}
    \renewcommand{\arraystretch}{1.1}
    \begin{table}[ht]
    \caption{ISE values for median estimators (multiplied by 1,000) at start, middle and end of simulated time frame. Smallest ISE values in each row are highlighted in bold.}
    \label{table:ISE_values}
    \centering 
        \begin{tabular}[t]{|c|c|c|c|c|c|c|c|c|}
        \hline
        Copula & Times & $\Bar{\lambda}_{QR}^*$ & $\Bar{\lambda}_{QR2}^*$ & $\Bar{\lambda}_{BP}^*$ & $\Bar{\lambda}_{BP2}^*$ & $\Bar{\lambda}_{GAM}^*$ & $\Bar{\lambda}_{GAM2}^*$ & $\Bar{\lambda}_{GAM3}^*$\\
        \hline
        Gaussian (Positive Correlation) & Start & \textbf{0.236} & 0.804 & 1.376 & 0.554 & 1.217 & 0.997 & 1.050\\
        \hline
        Gaussian (Positive Correlation) & Middle & 1.103 & 1.512 & 1.085 & 1.546 & 0.716 & 0.752 & \textbf{0.710}\\
        \hline
        Gaussian (Positive Correlation) & End & 0.012 & \textbf{0.006} & 0.338 & 0.167 & 0.089 & 0.090 & 0.088\\
        \hline
        Gaussian (Negative Correlation) & Start & 33314.799 & 37556.808 & 36091.579 & 38066.566 & \textbf{32039.496} & 34666.541 & 34858.391\\
        \hline
        Gaussian (Negative Correlation) & Middle & 81.758 & 79.520 & 81.964 & 76.083 & \textbf{62.132} & 65.816 & 66.826\\
        \hline
        Gaussian (Negative Correlation) & End & 0.158 & 1.468 & 1.163 & 1.776 & 0.130 & \textbf{0.083} & 0.107\\
        \hline
        Inverted Logistic & Start & 0.005 & \textbf{0.001} & 0.374 & 0.171 & 0.056 & 0.070 & 0.068\\
        \hline
        Inverted Logistic & Middle & \textbf{0.007} & 0.082 & 0.071 & 0.119 & 0.049 & 0.033 & 0.039\\
        \hline
        Inverted Logistic & End & 0.403 & 0.574 & 1.446 & \textbf{0.399} & 0.778 & 0.701 & 0.758\\
        \hline
        Inverted Husler-Reiss & Start & 5.667 & 4.020 & \textbf{1.417} & 3.394 & 6.430 & 2.141 & 2.296\\
        \hline
        Inverted Husler-Reiss & Middle & 0.007 & \textbf{0.002} & 0.008 & 0.056 & 0.014 & 0.055 & 0.056\\
        \hline
        Inverted Husler-Reiss & End & 0.015 & \textbf{0.001} & 0.352 & 0.081 & 0.019 & 0.033 & 0.034\\
        \hline
        Inverted Asymmetric Logistic & Start & 0.388 & 0.204 & 0.334 & \textbf{0.082} & 0.403 & 0.390 & 0.435\\
        \hline
        Inverted Asymmetric Logistic & Middle & \textbf{0.017} & 0.030 & 0.020 & 0.029 & 0.164 & 0.217 & 0.177\\
        \hline
        Inverted Asymmetric Logistic & End & 0.461 & 0.179 & 0.375 & \textbf{0.020} & 1.084 & 0.890 & 0.956\\
        \hline
        Copula of model (4.1) & Start & 0.054 & 0.013 & 0.250 & 0.041 & \textbf{0.005} & 0.006 & 0.006\\
        \hline
        Copula of model (4.1) & Middle & \textbf{0.034} & 0.076 & 0.129 & 0.204 & 0.079 & 0.070 & 0.075\\
        \hline
        Copula of model (4.1) & End & 6.423 & \textbf{1.158} & 5.959 & 1.925 & 1.175 & 1.625 & 1.503\\
        \hline
    \end{tabular}
\end{table}

\end{landscape}

We observe that the MISE and ISE values for the negatively correlated Gaussian copula at the start ($t=1$) and middle ($t=n/2$) of the interval are significantly larger than other values. This is due to the fact that for strongly negatively dependent data structures, the value of the true ADF tends towards infinity as $\rho$ approaches $-1$ for any ray $w \in (0,1)$, which is difficult to capture in practice. However, we note that while significant bias appears to exist in ADF estimates, we are still able to obtain accurate return curve estimates.

Overall, we conclude that the Bernstein polynomial and GAM-based estimators appear to perform best overall; while they may exhibit slightly more bias for certain copula copulas, this is counterbalanced with their lower variance, and hence lower MISE values. In the context of extreme value theory, it is important to select estimators that balance both variance and bias.    

\subsection{Non-stationary ADF estimates over time}
\begin{figure}[H]
    \centering
    \includegraphics[width=\textwidth]{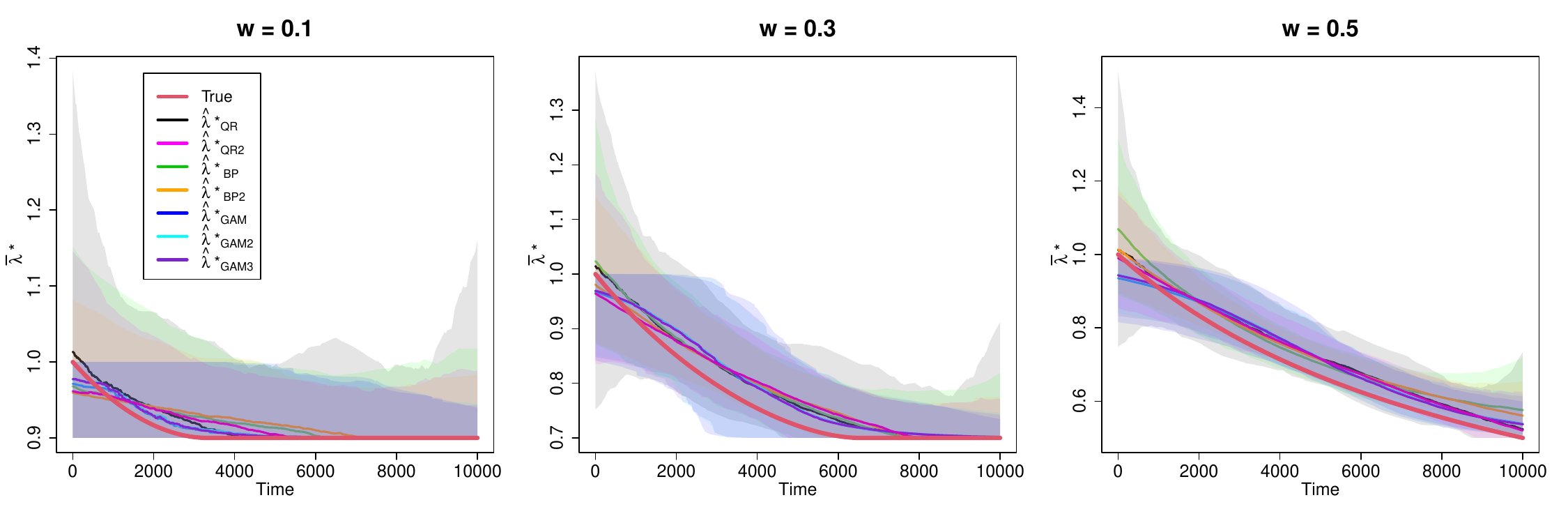}
    \caption{Non-stationary ADF estimates over time for the Gaussian copula with positive correlation.}
    \label{fig:bias_normal}
\end{figure}

\begin{figure}[H]
    \centering
    \includegraphics[width=\textwidth]{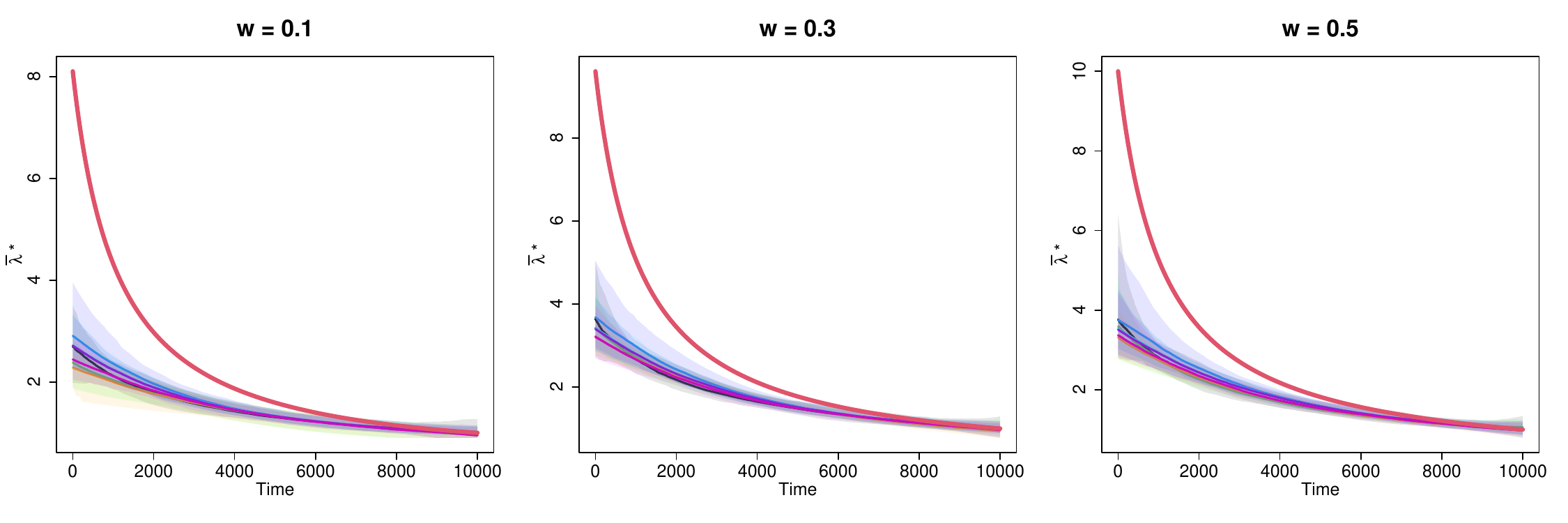}
    \caption{Non-stationary ADF estimates over time for the Gaussian copula with negative correlation. Legend is as in Figure \ref{fig:bias_normal}.}
    \label{fig:bias_normal2}
\end{figure}

\begin{figure}[H]
    \centering
    \includegraphics[width=\textwidth]{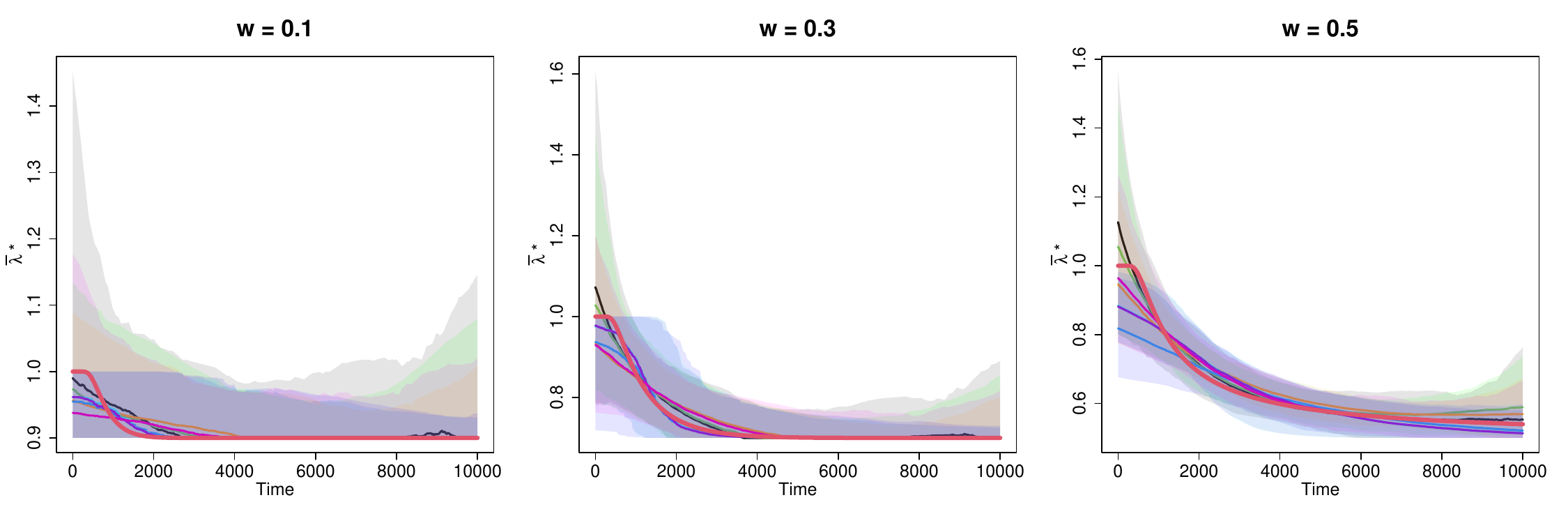}
    \caption{Non-stationary ADF estimates over time for the inverted H\"usler-Reiss copula.  Legend is as in Figure \ref{fig:bias_normal}.}
    \label{fig:bias_invhr}
\end{figure}

\begin{figure}[H]
    \centering
    \includegraphics[width=\textwidth]{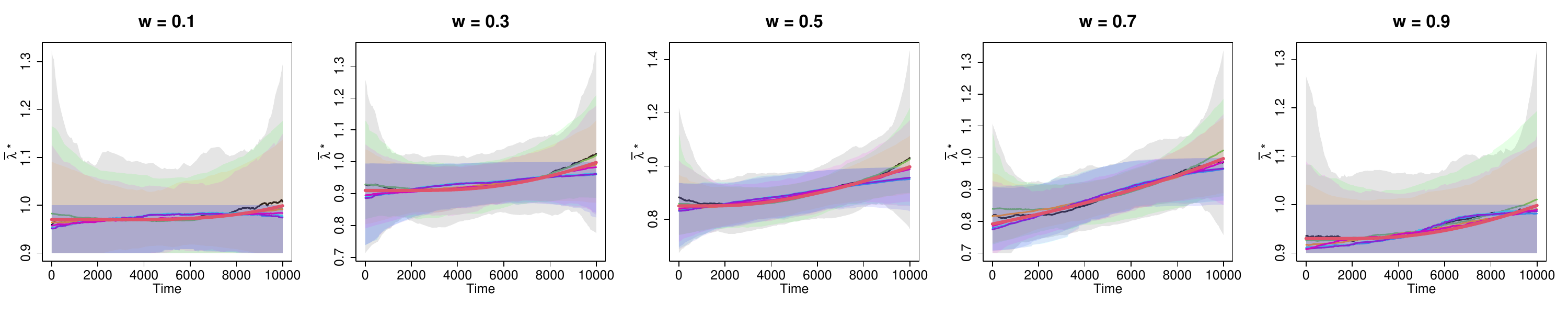}
    \caption{Non-stationary ADF estimates over time for the inverted asymmetric logistic copula. Legend is as in Figure \ref{fig:bias_normal}.}
    \label{fig:bias_invalog}
\end{figure}

\begin{figure}[H]
    \centering
    \includegraphics[width=\textwidth]{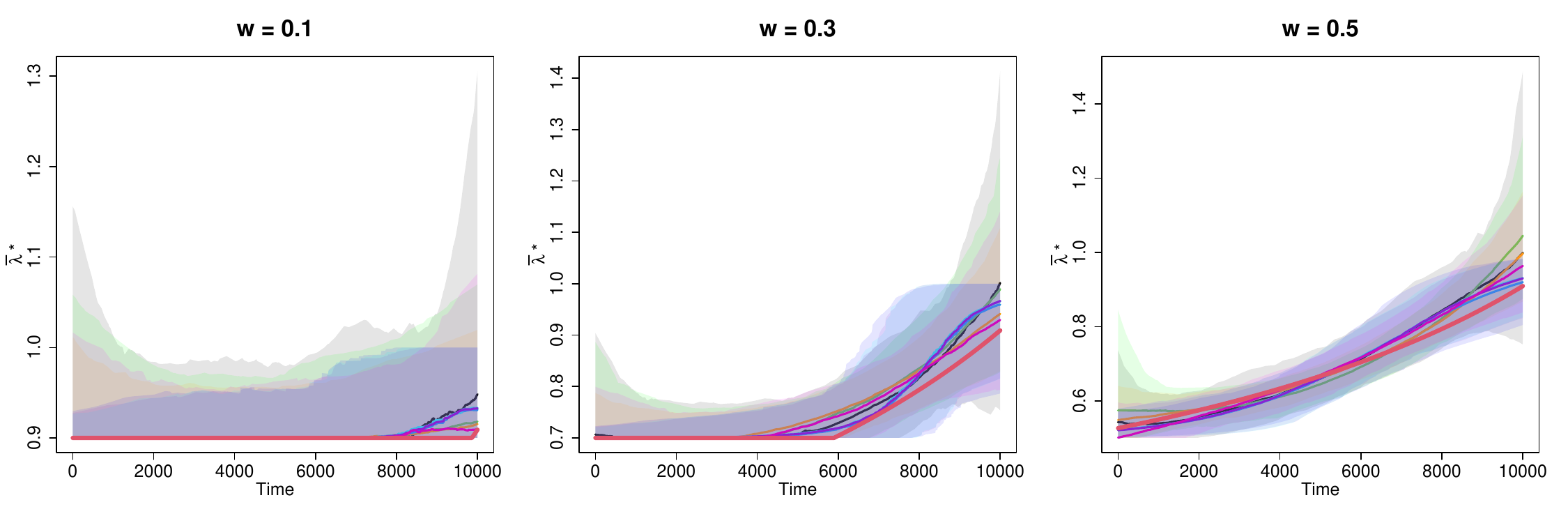}
    \caption{Non-stationary ADF estimates over time for the copula of model (4.1). Legend is as in Figure \ref{fig:bias_normal}.}
    \label{fig:bias_gauge}
\end{figure}

Figures \ref{fig:bias_normal} - \ref{fig:bias_gauge} illustrate median estimates, alongside $0.025$ and $0.975$ quantile estimates, of the ADF at fixed rays over time for each copula example, excluding the inverted logistic copula which has already been considered in the main text. The legend describing the line colours can be found in Figure 3 of the main article, with the coloured regions representing the area between the pointwise 0.025 and 0.975 quantiles. In each case, the rays $w = 0.1$, $w = 0.3$ and $w = 0.5$ have been considered. For the inverted asymmetric logistic copula, two additional rays ($w = 0.7$ and $w = 0.9$) have been considered to account for the asymmetry within this example.

\subsection{Non-stationary ADF estimates at fixed time points}
\begin{figure}[H]
    \centering
    \includegraphics[width=\textwidth]{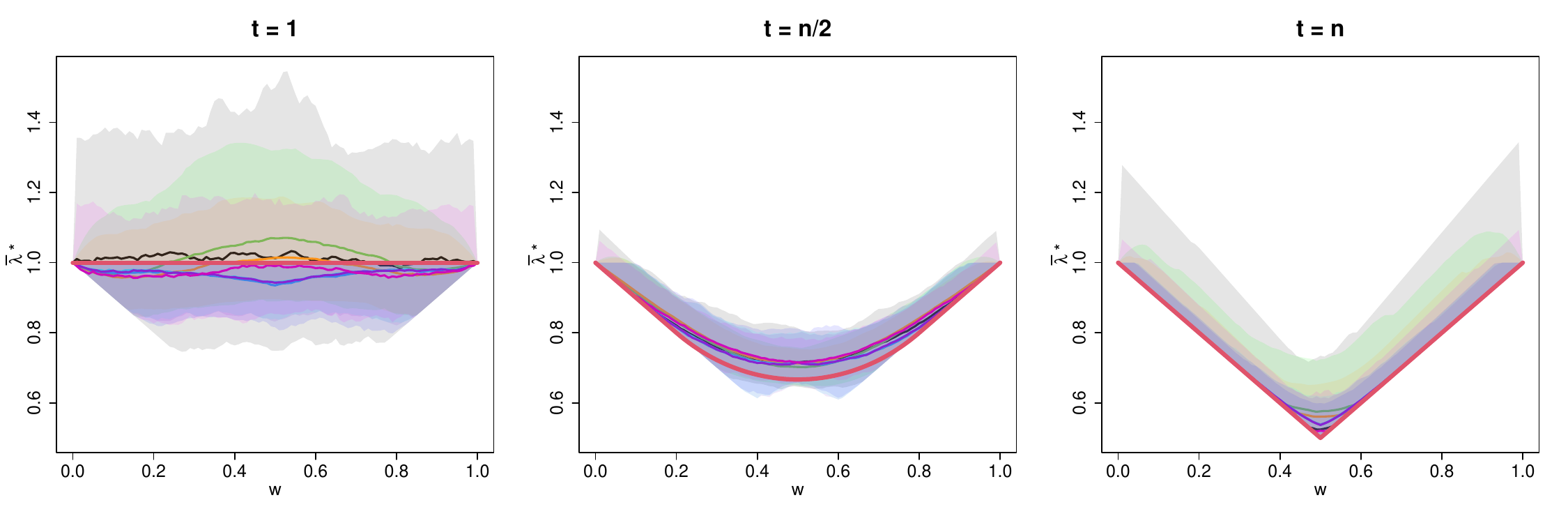}
    \caption{Non-stationary ADF estimates at three fixed time points for the Gaussian copula with positive correlation. Legend is as in Figure \ref{fig:bias_normal}.}
    \label{fig:uncertainty_normal}
\end{figure}

\begin{figure}[H]
    \centering
    \includegraphics[width=\textwidth]{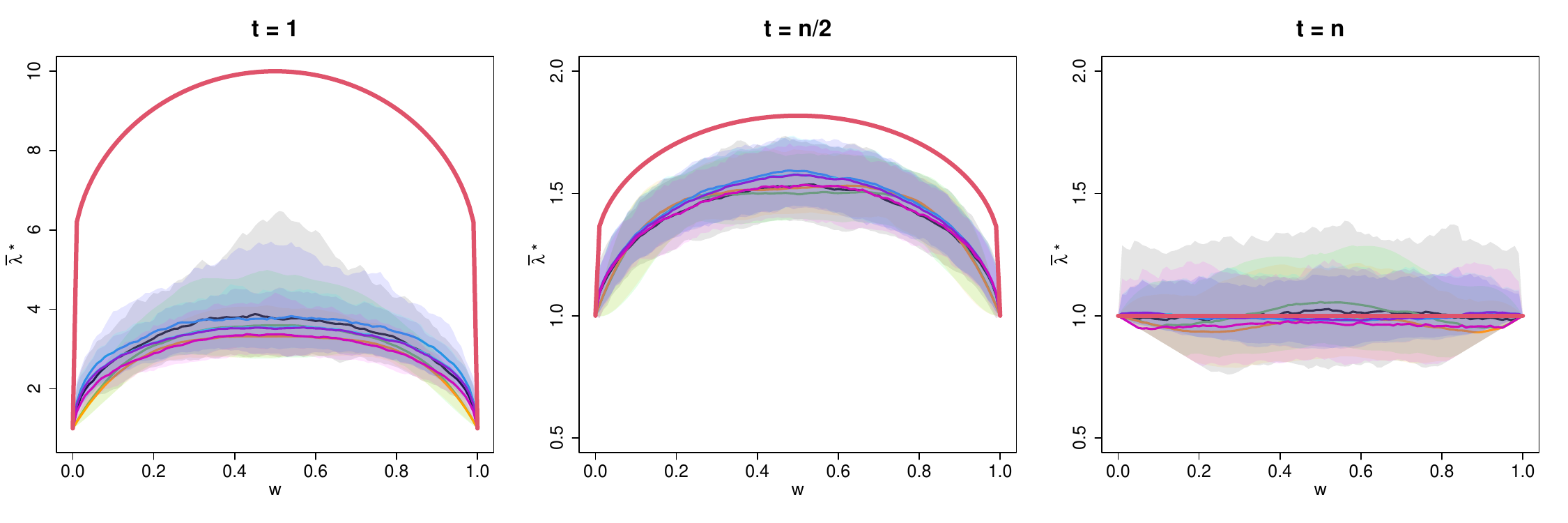}
    \caption{Non-stationary ADF estimates at three fixed time points for the Gaussian copula with negative correlation. Legend is as in Figure \ref{fig:bias_normal}.}
    \label{fig:uncertainty_normal2}
\end{figure}

\begin{figure}[H]
    \centering
    \includegraphics[width=\textwidth]{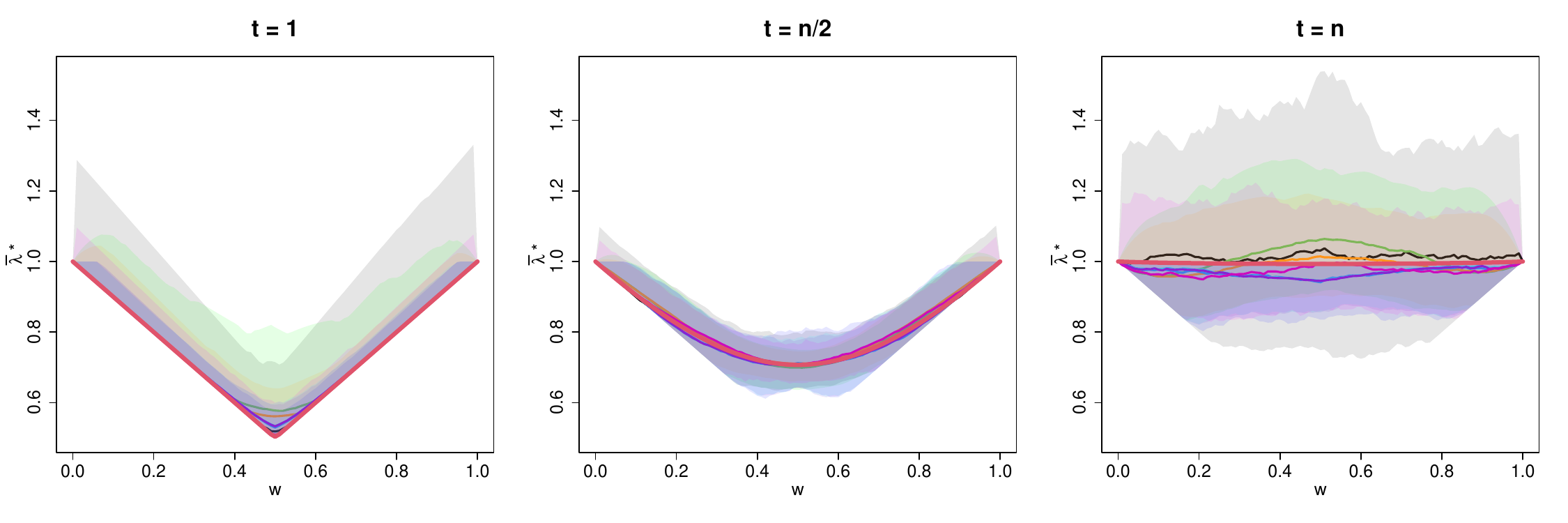}
    \caption{Non-stationary ADF estimates at three fixed time points for the inverted logistic copula. Legend is as in Figure \ref{fig:bias_normal}.}
    \label{fig:uncertainty_invlog}
\end{figure}

\begin{figure}[H]
    \centering
    \includegraphics[width=\textwidth]{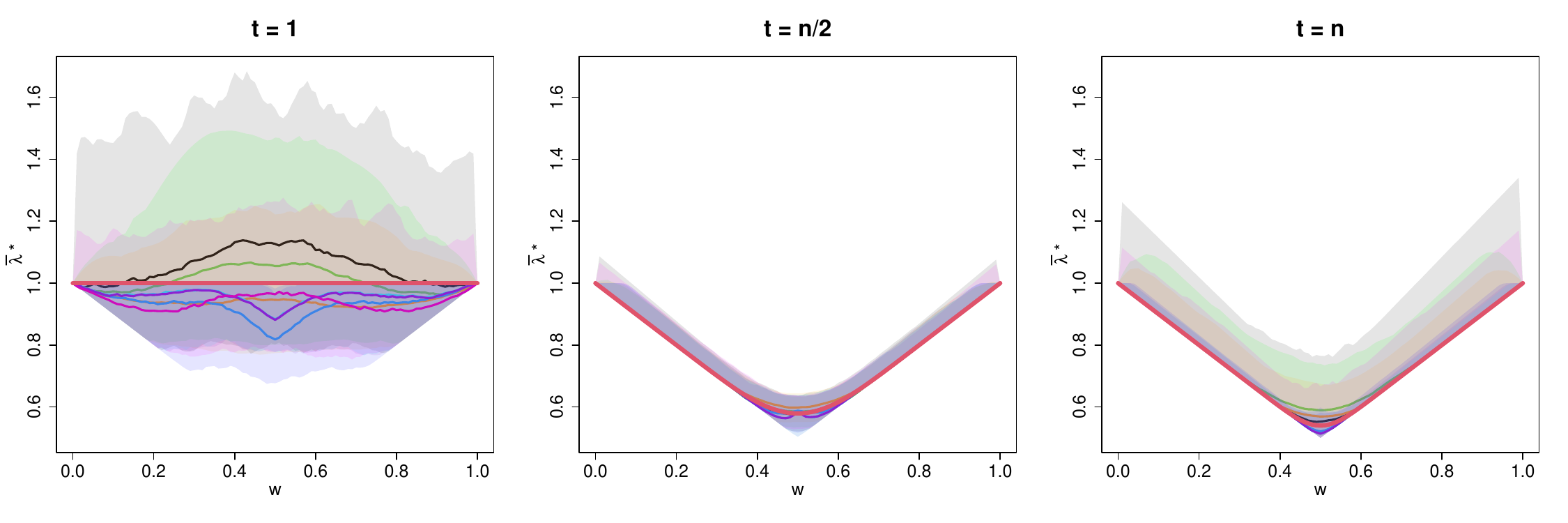}
    \caption{Non-stationary ADF estimates at three fixed time points for the inverted H\"usler-Reiss copula. Legend is as in Figure \ref{fig:bias_normal}.}
    \label{fig:uncertainty_invhr}
\end{figure}

\begin{figure}[H]
    \centering
    \includegraphics[width=\textwidth]{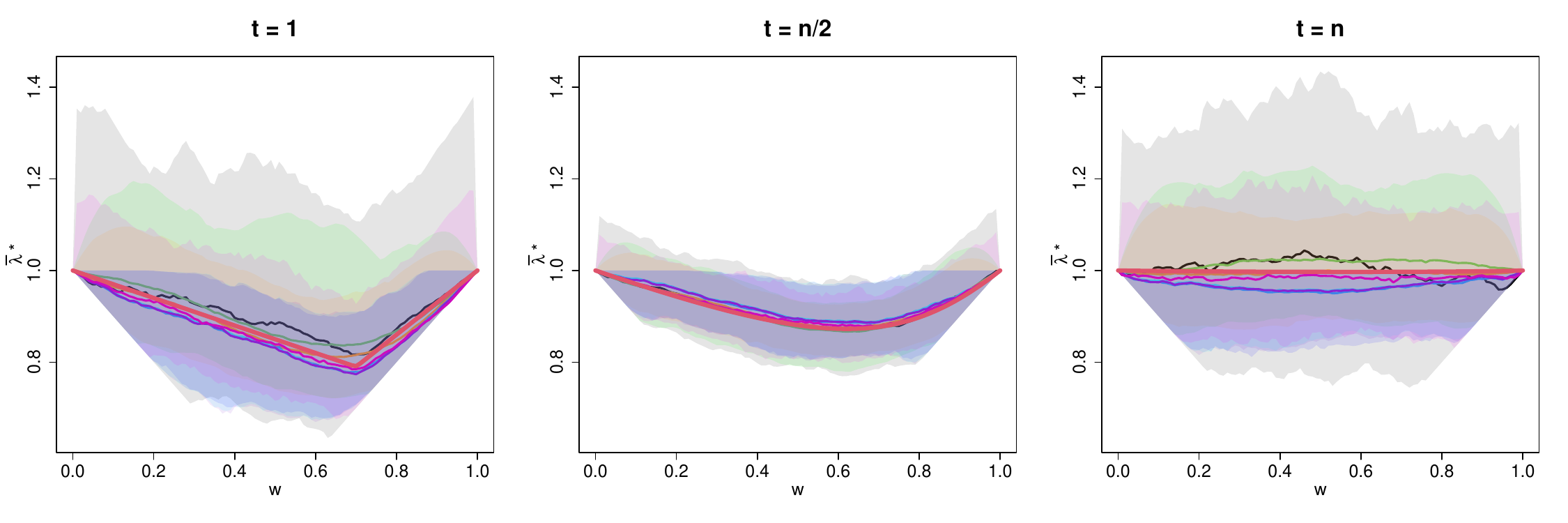}
    \caption{Non-stationary ADF estimates at three fixed time points for the inverted asymmetric logistic copula. Legend is as in Figure \ref{fig:bias_normal}.}
    \label{fig:uncertainty_invalog}
\end{figure}

\begin{figure}[H]
    \centering
    \includegraphics[width=\textwidth]{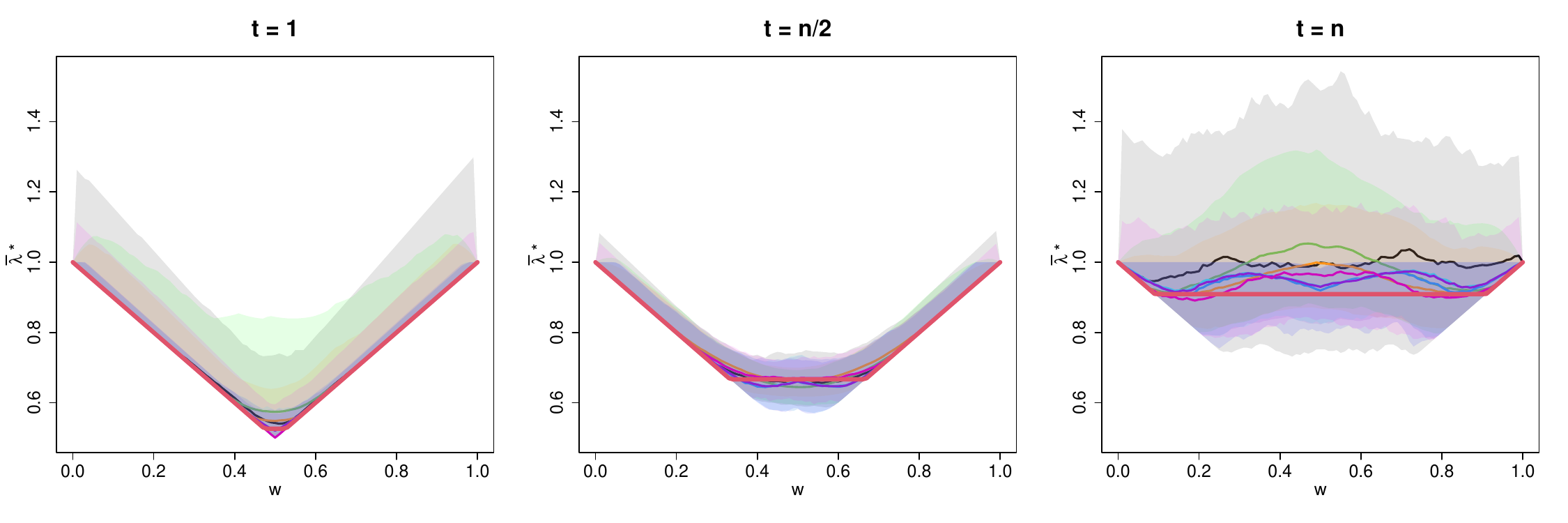}
    \caption{Non-stationary ADF estimates at three fixed time points for the copula of model (4.1). Legend is as in Figure \ref{fig:bias_normal}.}
    \label{fig:uncertainty_gauge}
\end{figure}

Figures \ref{fig:uncertainty_normal} - \ref{fig:uncertainty_gauge} illustrate median estimates, alongside $0.025$ and $0.975$ quantile estimates, of the ADF at three fixed time points ($t=1$, $t=n/2$ and $t=n$) over all rays $w \in [0,1]$ for each copula example. The legend describing the line colours can again be found in the main text, with the coloured regions representing the area between the pointwise 0.025 and 0.975 quantiles. 

\section{Additional simulation study results for a non-stationary dependence structure with two covariates}
\begin{figure}[H]
    \centering
    \includegraphics[width=.5\textwidth]{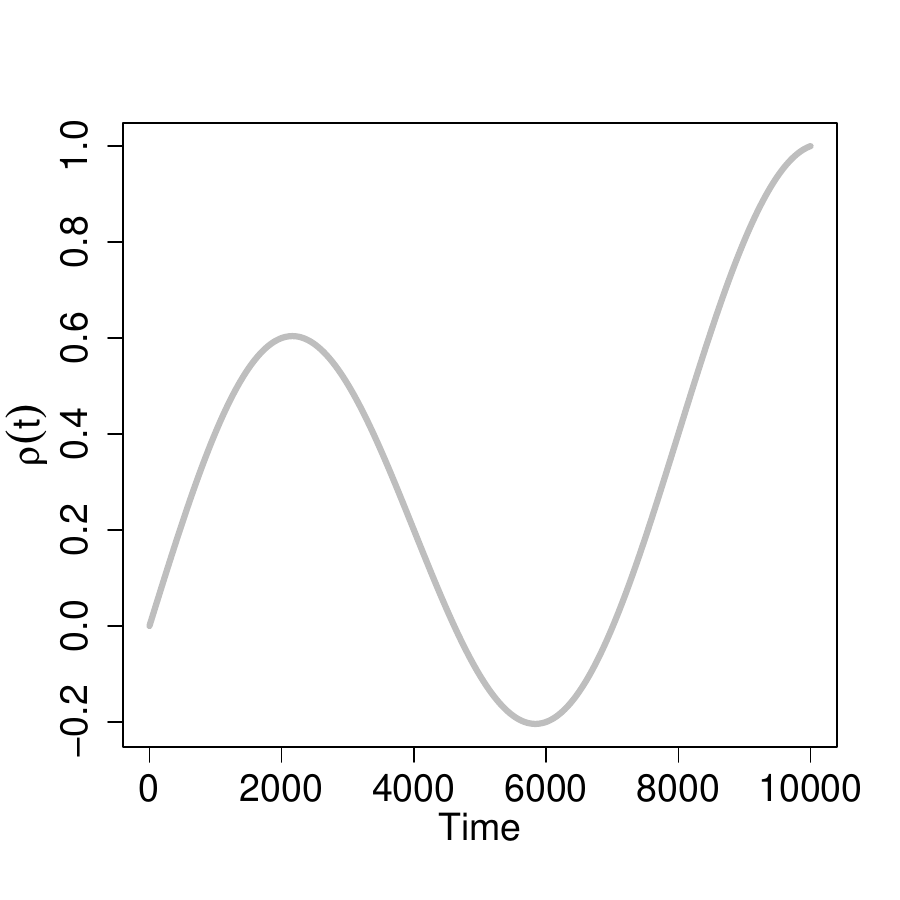}
    \caption{Plot of $\rho(t)$ over $t$ for the two covariate copula example.}
    \label{fig:rhos_twocov}
\end{figure}

Figure \ref{fig:rhos_twocov} illustrates the correlation coefficient function $\rho(t)$ over time for the two covariate copula example. One can observed the complex nature of the dependence trend for this case. 

Table \ref{table:MISE_values_twocov} gives MISE values for $\Bar{\lambda}_{BP2}^*$ and $\Bar{\lambda}_{GAM3}^*$, while Table \ref{table:ISE_values_twocov} gives the ISE values for the corresponding median estimators. As can be observed, in most cases, $\Bar{\lambda}_{BP2}^*$ outperforms $\Bar{\lambda}_{GAM3}^*$, suggesting the former estimator is better suited to capturing more complex dependence structures. 

\renewcommand{\arraystretch}{1.1}
\begin{table}[ht]
\caption{MISE values (multiplied by 1,000) at start, middle and end of simulated time frame. Smallest MISE values in each row are highlighted in bold.}
\label{table:MISE_values_twocov}
\centering 
    \begin{tabular}[t]{|c|c|c|c|c|c|c|c|c|}
    \hline
    Copula & Times & $\Bar{\lambda}_{BP2}^*$ & $\Bar{\lambda}_{GAM3}^*$\\
    \hline
    Gaussian (Two Covariate) & Start & \textbf{14.058} & 30.433\\
    \hline
    Gaussian (Two Covariate) & Middle & \textbf{7.920} & 8.757\\
    \hline
    Gaussian (Two Covariate) & End & 2.877 & \textbf{2.399}\\
    \hline
    \end{tabular}
\end{table}

\renewcommand{\arraystretch}{1.1}
\begin{table}[ht]
\caption{ISE values for median estimators (multiplied by 1,000) at start, middle and end of simulated time frame. Smallest ISE values in each row are highlighted in bold.}
\label{table:ISE_values_twocov}
\centering 
    \begin{tabular}[t]{|c|c|c|c|c|c|c|c|c|}
    \hline
    Copula & Times & $\Bar{\lambda}_{BP2}^*$ & $\Bar{\lambda}_{GAM3}^*$\\
    \hline
    Gaussian (Two Covariates) & Start & \textbf{2.853} & 10.538\\
    \hline
    Gaussian (Two Covariates) & Middle & 5.230 & \textbf{1.726}\\
    \hline
    Gaussian (Two Covariates) & End & \textbf{0.013} & 0.244\\
    \hline
    \end{tabular}
\end{table}

\begin{figure}[H]
    \centering
    \includegraphics[width=\textwidth]{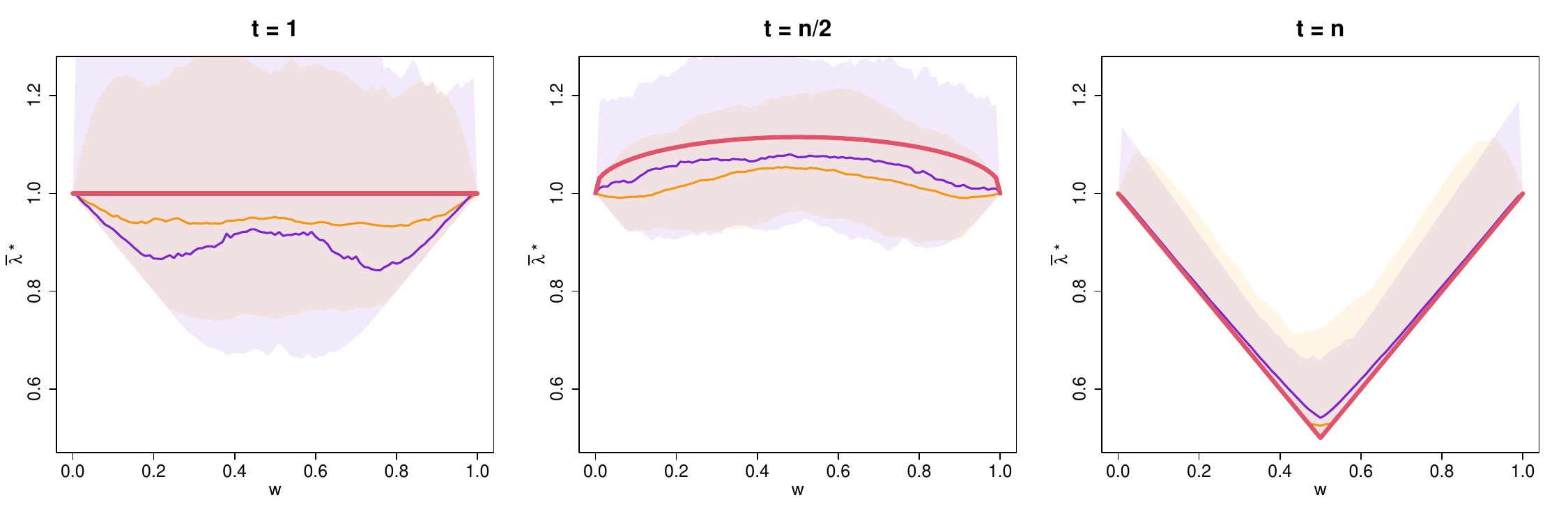}
    \caption{Non-stationary ADF estimates at three fixed time points for the two covariate Gaussian copula. Legend is as in Figure \ref{fig:bias_normal}.}
    \label{fig:uncertainty_twocov}
\end{figure}

Figure \ref{fig:uncertainty_twocov} illustrates median estimates, alongside $0.025$ and $0.975$ quantile estimates, of the ADF at three fixed time points ($t=1$, $t=n/2$ and $t=n$) over all rays $w \in [0,1]$ for the two covariate copula example. One can observe that $\Bar{\lambda}_{BP2}^*$ exhibits less variability than $\Bar{\lambda}_{GAM3}^*$. 


\section{Additional case study figures}

\subsection{Fitted pre-processing trend functions}
\begin{figure}[H]
    \centering
    \includegraphics[width=\textwidth]{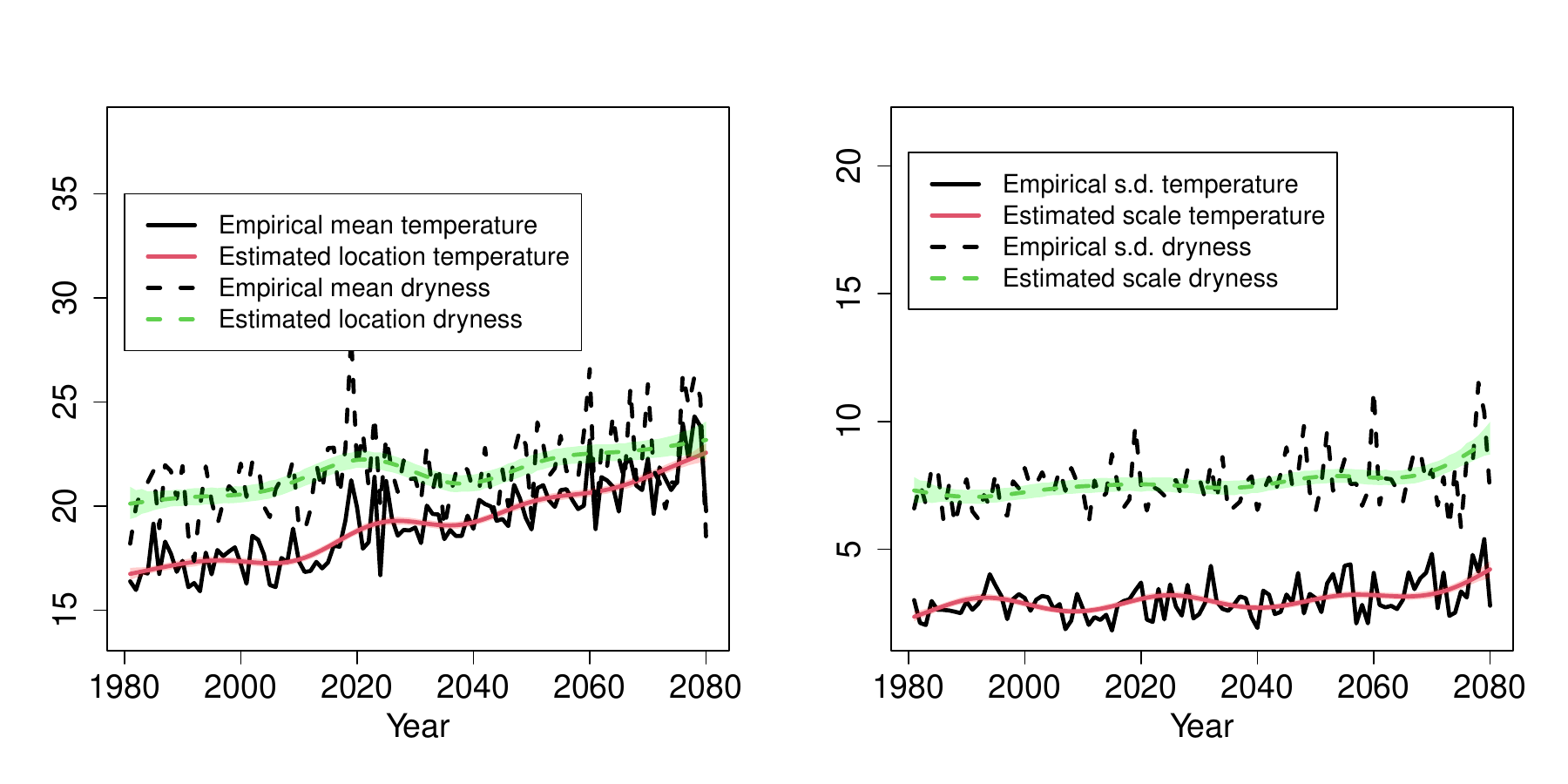}
    \caption{Comparison of estimated location and scale function values (red and green for temperature and dryness, respectively) with 95\% confidence intervals (shaded regions) against empirical mean and standard deviation estimates (black). For the fitted functions, the average value for a given year has been taken to ensure correspondence with the empirical values.}
    \label{fig:fitted_vs_true}
\end{figure}

Figure \ref{fig:fitted_vs_true} compares empirical estimates of the mean and standard deviations for both sets of projections against fitted location and scale functions, alongside estimated 95\% confidence regions. Empirical estimates are obtained using the data for fixed years over the observation period. The fitted location and scale functions are then averaged over each year and compared to the empirical estimates. One can observe very similar trends for both variables, indicating the pre-processing technique is accurately capturing the marginal non-stationary trends within the body of data. Note that the estimated confidence regions are likely to be an under-representation of the true uncertainty due to the fact we have assumed independence between observations when fitting the GAMs.

\subsection{Estimated rate parameters}
\begin{figure}[H]
    \centering
    \includegraphics[width=\textwidth]{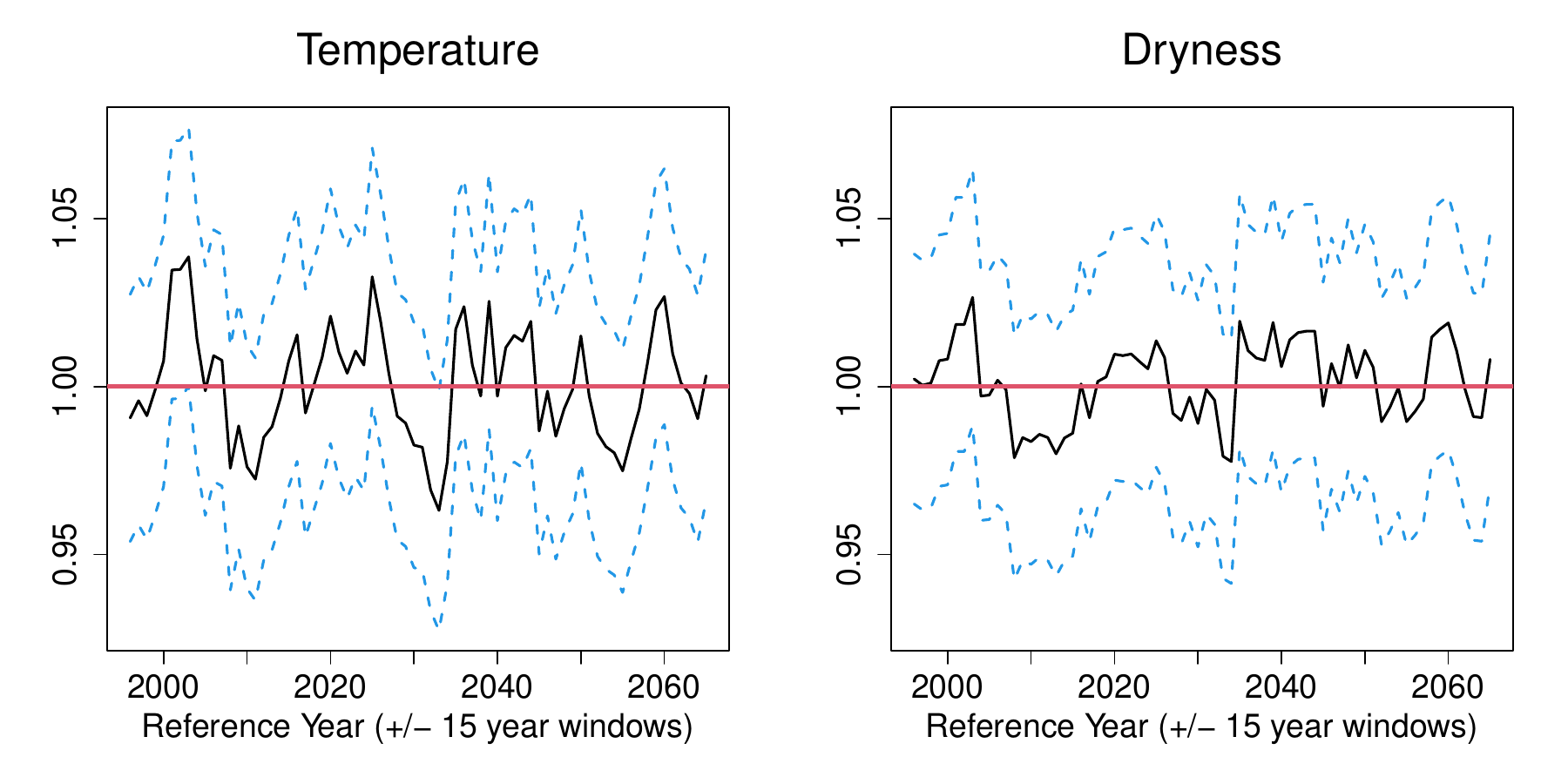}
    \caption{Estimated exponential rate parameters (black) with 95\% pointwise confidence intervals (dotted blue) over the time period. The target rate parameter is given in red.}
    \label{fig:rate_paras}
\end{figure}

Figure \ref{fig:rate_paras} illustrates exponential rate parameter estimates for the pre-processed data, alongside 95\% pointwise confidence intervals, for $\pm 15$ year rolling windows over the observation period. As can be observed, the rate parameter estimates remain approximately constant at one throughout the entire observation period, suggesting a successful transformation to exponential margins.  

\end{document}